\documentclass[11pt, dvipsnames]{scrartcl}
\usepackage{mathtools}
\usepackage{cmll}
\usepackage{mdframed}
\usepackage{subcaption}
\usepackage{tikz}
\usetikzlibrary{calc, shapes, backgrounds, automata, positioning, arrows}
\usepackage{stmaryrd}
\usepackage{hyperref}
\usepackage{ebproof}
\usepackage{proof}
\usepackage{amssymb}
\usepackage{amsthm}
\usepackage[title]{appendix}
\usepackage{glossaries}
\usepackage{xparse}
\usepackage{amsthm,amssymb,amsmath}
\usepackage[most]{tcolorbox}
\newcommand{\nat}{\mathbf{N}}

\newcommand*{\topb}{\mathsf{T}}
\newcommand*{\botb}{\rotatebox[origin=c]{-180}{$\topb$}}

\DeclareMathOperator{\downtens}{\downarrow\!}
\DeclareMathOperator{\uppar}{\uparrow\!}

\newcommand{\dgstar}[2]{[\frac{#1}{#2}]}

\newcommand{\eqq}{\stackrel{?}{=}}

\newcommand{\exec}{\mathtt{Ex}}

\newcommand{\iostar}[2]{\left[\frac{#1}{#2}\right]}

\definecolor{bgstellar}{rgb}{0.9,0.9,0.9}

\newcommand{\tstar}[4][20]{
    \foreach \x [count=\i from 0, count=\j from -1] in #4
    {
        \ifnum \i=0
            \node[draw=bgstellar, anchor=west, minimum height=#1pt, fill=bgstellar] (#2/0) at #3 {\x;};
        \else
            \node[draw=bgstellar, anchor=west, minimum height=#1pt, fill=bgstellar] (#2/\i) at (#2/\j.east) {\x;};
        \fi
    }
}

\newcommand{\unistar}[4][20]{\tstar[#1]{#2}{#3}{{{#4}}}}
\newcommand{\bistar}[5][20]{\tstar[#1]{#2}{#3}{{{#4},{#5}}}}

\newcommand{\addr}{\mathtt{addr}}

\newcommand{\fu}{\text{\begin{CJK}{UTF8}{}\begin{Japanese}\textbf{フ}\end{Japanese}\end{CJK}}}
\newcommand{\wo}{\text{\begin{CJK}{UTF8}{}\begin{Japanese}\textbf{ヲ}\end{Japanese}\end{CJK}}}

\newcommand{\wang}[4]{
    \begin{tikzpicture}
        \draw [black,fill=#4] (0,0)--(0.5,0.5)--(0,1)--cycle;
        \draw [black,fill=#3] (0,0)--(0.5,0.5)--(1,0)--cycle;
        \draw [black,fill=#2] (1,1)--(0.5,0.5)--(1,0)--cycle;
        \draw [black,fill=#1] (1,1)--(0.5,0.5)--(0,1)--cycle;
    \end{tikzpicture}
}

\NewDocumentCommand{\domino}{mm}{
\begin{tikzpicture}[x=2em,y=2em,radius=0.1]
\draw[rounded corners=0.5,line hidden] (0,0) rectangle (1,2);

\draw[line hidden] (0,1) -- (1,1);
\ifodd#1
  \fill[dot hidden] (0.5,1.5) circle;
\fi
\ifnum#1>1
   \fill[dot hidden] (0.2,1.2) circle;
   \fill[dot hidden] (0.8,1.8) circle;
   \ifnum#1>3
      \fill[dot hidden] (0.8,1.2) circle;
      \fill[dot hidden] (0.2,1.8) circle;
   \fi
   \ifnum#1>5
      \fill[dot hidden] (0.2,1.5) circle;
      \fill[dot hidden] (0.8,1.5) circle;
   \fi
\fi
\ifodd#2
   \fill[dot hidden] (0.5,0.5) circle;
\fi
\ifnum#2>1
   \fill[dot hidden] (0.2,0.2) circle;
   \fill[dot hidden] (0.8,0.8) circle;
   \ifnum#2>3
      \fill[dot hidden] (0.8,0.2) circle;
      \fill[dot hidden] (0.2,0.8) circle;
   \fi
   \ifnum#2>5
      \fill[dot hidden] (0.2,0.5) circle;
      \fill[dot hidden] (0.8,0.5) circle;
   \fi
\fi
\end{tikzpicture}
}

\newcommand{\gc}{\mathtt{g}}
\newcommand{\ccol}[1]{{\color{Bittersweet}#1}}
\newcommand{\tcol}[1]{{\color{MidnightBlue}#1}}
\newcommand{\ulocus}[2]{p_{#1}(#2)}
\newcommand{\clocus}[2]{\ccol{+c.\ulocus{#1}{#2}}}
\newcommand{\tlocus}[3][+]{\tcol{#1 t.\ulocus{#2}{#3}}}

\newcommand{\vcut}[2]{\ccol{-c.p_{#1}(#2)}}
\newcommand{\qray}[2][]{\ifthenelse{\equal{#1}{}}{p_{#2}(\gc \cdot x)}{\ccol{#1c.p_{#2}(\gc \cdot x)}}}
\newcommand{\gtens}[2]{
    [\qray[-]{#1}, \qray[-]{#2}, \qray[+]{#1 \otimes #2}]
}
\newcommand{\gparr}[2]{
    [\qray[-]{#1}] +
    [\qray[-]{#2}, \qray[+]{#1 \parr #2}]
}
\newcommand{\gparl}[2]{
    [\qray[-]{#1}, \qray[+]{#1 \parr #2}] +
    [\qray[-]{#2}]
}

\newcommand{\gconc}[1]{[\qray[-]{#1}, p_{#1}(x)]}

\newcommand{\dgloc}[3]{\dgstar{\tcol{-t.p_{#1}(#2)}}{\ccol{+c.q_{#3}(x)}}}
\newcommand{\dgtens}[3]{
    \dgstar{\ccol{-c.q_{#1}(x)} \quad \ccol{-c.q_{#2}(x)}}
    {\ccol{+c.q_{#3}(x)}}
}

\newcommand{\dgparl}[3]{
    \dgstar{\ccol{-c.q_{#1}(x)}}{\ccol{+c.q_{#3}(x)}} +
    \dgstar{\ccol{-c.q_{#2}(x)}}{}
}
\newcommand{\dgconc}[1]{\dgstar{\ccol{-c.q_{#1}(x)}}{p_{#1}(x)}}

\newcommand{\gaxex}[2]{[\tcol{-t.\addr_{#1}(#2)}, \ccol{+c.q_{#2}(x \cdot y)}]}

\usepackage{CJKutf8}

\newenvironment{Japanese}{%
    \CJKfamily{min}%
    \CJKtilde
    \CJKnospace}{}

\newtheorem{example}{Example}
\newtheorem{definition}{Definition}
\newtheorem{proposition}{Proposition}

\tikzset{
    dot/.style = {fill=black, circle, outer sep=2.5pt, inner sep=1pt},
    every loop/.style={},
    dot hidden/.style={},
    line hidden/.style={},
    dot colour/.style={dot hidden/.append style={color=#1}},
    dot colour/.default=black,
    line colour/.style={line hidden/.append style={color=#1}},
    line colour/.default=black
}

\begin{document}

\title{A gentle introduction to Girard's Transcendental Syntax for the linear logician}
\subtitle{Version 7}
\author{Boris Eng}
\date{}
\maketitle

\begin{abstract}
Technically speaking, the transcendental syntax is about designing logics with a computational foundation. It suggests a new framework for proof theory where logic (proofs, formulas, truth, ...) is no more primitive but computation is. All the logical entities and activities will be presented as formatting/structuring on a given model of computation which should be as general, simple and natural as possible. The selected ground for logic in the transcendental syntax is a model of computation I call "stellar resolution" which is basically a logic-free reformulation of Robinson's first-order clausal resolution with a dynamics related to tile systems. An initial goal of the transcendental syntax is to retrieve linear logic from this new framework. In particular, this model naturally encodes cut-elimination for proof-structures. By using an idea of ``interactive typing'' reminiscent of realisability theory, it is possible to design formulas/types generalising the connectives of linear logic. Thanks to interactive typing, we are able to reach a semantic-free space where correctness criteria are seen as tests (as in unit testing or model checking) certifying logical correctness, thus allowing an effective use of logical entities.
\end{abstract}

A friendly glossary of the transcendental syntax and of Girard's terminology is presented at the end of this paper. Please contact me by e-mail if this introduction is not gentle enough in your opinion.

\bigskip
\textbf{Prerequisites:} linear logic, sequent calculus, proof-nets, correctness criteria, term unification, first-order resolution.

\tableofcontents

\section{Introduction}

\subsection{History and scientific context}

The transcendental syntax is the direct successor of Girard's \emph{Geometry of Interaction} (GoI) \cite{goi0, goi1, goi2, goi3, goi4, goi5}. The initial idea of the GoI was to study cut-elimination in a purely computational and mathematical fashion. For instance, in the case of MLL, Girard remarked \cite{girard1987multiplicatives} that it was sufficient to represent proofs as permutations on a finite subset $S \subseteq \nat$ of natural numbers (representing logical atoms) and cut-elimination as a procedure of interaction between permutations\footnote{This idea has been simplified with graph theory by Seiller \cite{seiller2012interaction}}. Then an extension to exponentials (handling of erasure and duplication of formulas), would naturally require infinitely many natural numbers for the potentially infinite copies of logical atoms. This is why Girard introduced interpretations with operator algebras \cite{goi1}. The next articles of GoI are extensions of this idea to full linear logic and more. The current most complete historical introduction on the GoI is probably Seiller's thesis \cite{seillerPHD} (written in French).

Unsatisfied with the infinite and complicated objects obtained, Girard introduced the transcendental syntax as a successor which would use simple and finite combinatorics, by extending a simplification of an idea coming from the third paper of GoI \cite{goi3}. Philosophical motivations such as the wish for the sufficient conditions for a finite and tractable reasoning in a chaotic semantic-free space also appeared along this new project. Moreover, inspired by the GoI, logic would not be primitive anymore: instead, we should start from a very general idea of computation and then a logic would be designed for this model. In some sense, it is opposite to the practice of the proof-program correspondence: instead of taking a logic and turning it into a type system for a new programming language, we design a logic for a programming language speaking about its computational behaviour. In the case of linear logic, I personally like to see it as a sort of \emph{reverse engineering} of logic where logic would be \emph{retrieved} instead of \emph{pre-defined}.

Note that although the GoI has a quite rich history, it was a rather isolated series of works. In the literature, the term ``Geometry of Interaction" usually refers to a simplification of Girard's first GoI \cite{goi1} suggested by Danos and Regnier \cite{danos1995proof} which led to the idea of \emph{token machine} \cite{danos1999reversible, asperti1995paths} used for rewriting-free evaluations of $\lambda$-terms and which also inspired categorical semantics of linear logic \cite{abramsky2002geometry, haghverdi2006categorical}.

\subsection{Philosophical motivations}

In this section, I assume that the reader is already (even vaguely) aware of Girard's intuition and conception of logic without necessarily understanding them.

A quite old and recurrent idea of Girard is the idea of \emph{blind spot}: the idea that our understanding of logic is biased and only an approximation of a bigger picture we do not fully understand. Indeed, the whole study of logic usually lies on primitive definitions and preconception of what a proof or a formula should be. Reductionist approaches of logic are traditions in mathematical logic: the mechanisms of these given concepts are explored using mathematics.

In the GoI, Girard tried the opposite approach of founding logic over very general computational entities without any logical meaning. By using an idea of \emph{interactive typing} (also appearing in ludics \cite{girard2001locus} or classical realisability \cite{krivine2009interactive}), it is possible to speak logic in a semantic-free space since formulas are constructed as grouping of computational entities instead of simply being predefined syntactic labels.

The philosophical problem that the transcendental syntax aims to solve and which was not present in the GoI is that in this chaotic semantic-free space, finite and tractable reasoning is not possible anymore: it is theoretically impossible to verify the logical correctness of even the simplest things (such as $\forall X.X\Rightarrow X$) for reasons we will see in this paper. The idea is to take inspiration from the correctness criteria of proof-net theory as a sufficient and finite way to assert logical correctness.

In the transcendental syntax writing a simple and naive proof in sequent calculus is not so innocent: proof-nets theory shows that when doing so, we implicitly manipulate computational objects holding a certificate asserting that the computational object is logically correct. Correct proofs are like individuals having a VIP card but we forget that there exist some other individuals not having it. What Girard claims is that these forgotten individuals are essential in the understanding of the mechanisms of logic. An ambition of the transcendental syntax is to make these implicit assumptions computationally explicit and to "put everything on the table, even the table".

In terms of linear logic, this idea is materialised by the correctness criterion of linear logic: a computational object (proof-structure) can be tested in order to tell whether or not it is logically correct (that it corresponds to a sequent calculus proof). In the transcendental syntax, this idea is generalised beyond proof-structures: any computational object is a ``proof'' of something when it is accompanied with some tests it satisfies. The idea of logical correctness then becomes relative/subjective. It means that what we mean by ``passing the tests'' is up to us, depending on what we would like to obtain (for instance, a specific computational behaviour), exactly as we would do when checking program specifications in model checking for instance.

Girard's philosophical motivation for the transcendental syntax is to establish a whole new architecture for logic which would be free of any logical preconception but also explain the whole logical activity. In this idea, logic is a formatting of computation: everything starts from a chosen very general, simple and natural model of computation. This is what Girard calls the \emph{analytics}, in reference to Kant. We require that the model includes reducible objects which can be evaluated (what Girard calls \emph{performance}) to an irreducible object (Girard's \emph{constat}). Both are separated by \emph{indecidability}: the fact that an reducible object cannot always be reduced to its potential results (infinite loop may appear).

Once we have a computational ground, logic is a subjective way to provide a meaning to the computational objects, what Girard calls \emph{synthetics}. There are two ways to make a computation meaningful. First, defining an arbitrary finite set of tests (as we would do in programming) and see if the object of interest passes these tests. Interestingly, these tests are also encoded as objects of the same kind as the tested and are part of the type theory of transcendental syntax instead of being external objects (like how tests in programming are also programs). This is Girard's \emph{usine} (factory in English). Another way to provide meaning is to classify objects depending on how they interact with other ones by the interactive typing mentioned above. This is Girard's \emph{usage} (use in English). Girard claims that the factory and the use are separated by \emph{Gödel's incompleteness theorems} because tests (our arbitrary definitions of proofs and formulas) cannot always perfectly guarantee the use of logical definitions (the capture of classical truth for instance). The tests are either too strict (and missing some use cases -- or some true formulas becoming unprovable) or too permissive (contradictions appear).

Notice that this architecture is founded on \emph{fatalities} and the idea of \emph{explicitation} appearing in both computation and logic. Hence, this suggestion of foundations for logic has a rather solid ground.

\subsection{Transcendental Syntax as a toolbox for computer science}

The idea of capturing computational properties of behaviour by formulas is not new to computer science. Formulas are convenient ways to reason about the regularity of computational phenomena by only manipulating syntactic labels. The transcendental syntax generalises this practice of formulas in a larger and more convenient framework.

\begin{itemize}
    \item In \emph{descriptive complexity} \cite{immerman2012descriptive}, complexity classes such as P and NP are captured by fragments of a logic and decision problems are characterised by formulas. In our case, we should be able to design atypical formulas existing outside a predefined logical system in order to describe computation in a more fine-grained way.
    \item In \emph{model checking}, we are interested in reasoning about properties of a model of computation representing a real system. We will see that in the model of computation we consider, it is possible to naturally encode various classes of circuits, state machines and tile systems but also to design formulas speaking about them. From these formulas (corresponding to specifications), tests can be constructed in order to check if our representation satisfies a specific specification.
    \item In \emph{unit testing}, a program is tested against a finite of tests in order to guarantee a specific computational behaviour and to certify a partial correctness. In the transcendental syntax, these tests can also be typed and unit testing applies to any model of computation we can express.
\end{itemize}

\section{Stellar Resolution: query-free concurrent first-order resolution}

The stellar resolution is the computational ground for logic selected by Girard. In fact, this is not a necessary choice at all. Other choices can be made but this one is especially convenient for linear logic or computable functions. It comes from a generalisation of flows (a model of computation appearing in the third article of GoI \cite{goi3}).

We use the name "stellar" for Girard's terminology of \emph{stars and constellations} and "resolution" for its similarities with other resolution-based models \cite{kowalski1975proof, sickel1976search}.

For pedagogical purposes, we describe our model of computation as a generalisation of tiling models and show that it behaves as a sort of concurrent, query-free and alogical logic programming language.

\subsection{Tile systems}

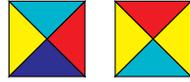
\begin{description}
    \item[Wang tiles \cite{wang}] $\wang{Turquoise}{Red}{Blue}{Yellow}\quad \wang{Red}{Yellow}{Turquoise}{Yellow}$ We consider bricks with four faces for the four directions in the plane $\mathbf{Z}^2$. We can connect two opposite faces of two bricks when they have the same colour. We are usually interested in constructing tilings by reusing the tiles as many times as possible (hence, producing maximal constructions). This is a well-known model and a good introduction to tiling-based computation.
    \item[Flexible tiles \cite{jonoska2006flexible, jonoska2005computational}]
    \begin{tikzpicture}
        \node[circle,draw=black] at (0, 0) (c1) {$t_1$};
        \node[dot, label={180:$h_1$}] at (-1, 0) (b1) {};
        \node[dot, label={90:$h_2$}] at (1, 0) (b2) {};
        \node[dot, label={180:$h_3$}] at (0, -1) (b3) {};
        \draw (c1) to[in=-30, out=150] (b1);
        \draw (c1) to[out=-30, in=150] (b2);
        \draw (c1) to[out=220, in=30] (b3);

        \node[circle,draw=black] at (5, 0) (c2) {$t_2$};
        \node[dot, label={90:$\theta(h_2)$}] at (4, 0) (a1) {};
        \node[dot, label={0:$h_4$}] at (6, 0) (a2) {};
        \draw (c2) to[in=-30, out=150] (a1);
        \draw (c2) to[out=-30, in=150] (a2);
        \draw[dashed] (b2) to[out=-30, in=150] (a1);
    \end{tikzpicture} This is a less common but still interesting model of computation. Instead of imposing planarity to tilings, we can allow tiles to have flexible sides selected from a set of \emph{sticky-ends types} $H$. We connect the sticky-ends w.r.t. an involution $\theta$ defining complementarity. Surprisingly, this model is able to simulate rigid tiles such as Wang tiles \cite{jonoska2006flexible}.
\end{description}

\subsection{Stars and constellations}

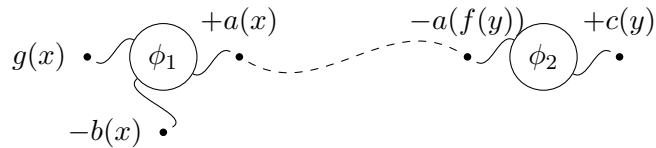
\begin{center}
\begin{tikzpicture}
    \node[circle,draw=black] at (0, 0) (c1) {$\phi_1$};
    \node[dot, label={180:$g(x)$}] at (-1, 0) (b1) {};
    \node[dot, label={90:$+a(x)$}] at (1, 0) (b2) {};
    \node[dot, label={180:$-b(x)$}] at (0, -1) (b3) {};
    \draw (c1) to[in=-30, out=150] (b1);
    \draw (c1) to[out=-30, in=150] (b2);
    \draw (c1) to[out=220, in=30] (b3);

    \node[circle,draw=black] at (5, 0) (c2) {$\phi_2$};
    \node[dot, label={90:$-a(f(y))$}] at (4, 0) (a1) {};
    \node[dot, label={90:$+c(y)$}] at (6, 0) (a2) {};
    \draw (c2) to[in=-30, out=150] (a1);
    \draw (c2) to[out=-30, in=150] (a2);

    \draw[dashed] (b2) to[out=-30, in=150] (a1);
\end{tikzpicture}
\end{center}

The stellar resolution (sketched in this section) can be seen as a flexible tiling model with terms called \emph{rays} as sticky-ends. A ray can either be polarised with a head symbol called \emph{colour} (e.g. $+a(x)$ or $-a(f(y))$ where $a$ is a colour and $t$ is any unpolarised term) or unpolarised (e.g. $x$ or $g(x)$).

A \emph{star} (tile) is a finite multiset of rays $\phi = [r_1, ..., r_n]$ and a \emph{constellation} $\Phi = \phi_1 + ... + \phi_m$ (tile set / program) is a (potentially infinite\footnote{Finite constellations are sufficient when speaking about logic.}) multiset of stars. We consider stars to be equivalent up to renaming and no two stars within a constellation share variables: these variables are local in the sense of programming. The empty star is written $[]$.

By using occurrences of stars from a given constellation, we can connect them along their matchable rays of opposite polarity in order to construct tilings called \emph{diagrams}.

We get the following correspondence of terminology:

\begin{center}
\begin{tabular}{|c|c|c|}
    \hline
    \textbf{Tile systems} & \textbf{Stellar Resolution} \\
    \hline
    Sticky-end & Ray $r = +a(t), -a(t), t$ \\
    \hline
    Tile & Star $\phi = [r_1, ..., r_n]$ \\
    \hline
    Tile set & Constellation $\Phi = \phi_1+...+\phi_m$ \\
    \hline
    Tiling & Diagram \\
    \hline
\end{tabular}
\end{center}

\begin{example}
\label{ex:logicprograms}
We encode two typical logic programs as constellations. We write $s^n$ for $n$ applications of the symbol $s$ in order to encode natural numbers. A colour corresponds to a predicate and the polarity represents the distinction input/output or hypothesis/conclusion. An unpolarised ray cannot interact with other rays.

\begin{verbatim}
add(0, y, y).
add(s(x), y, s(z)) :- add(x, y, z).
?add(s^n(0), s^m(0), r).
\end{verbatim}
\[\Phi_\nat^{n,m} = [+add(0, y, y)]+[-add(x, y, z), +add(s(x), y, s(z))]+[-add(s^n(0), s^m(0), r), r]\]

\bigskip
\begin{verbatim}
parent(d, j). parent(z, d). parent(s, z). parent(n, s).
ancestor(x, y) :- parent(x, y).
ancestor(x, z) :- parent(x, y), ancestor(y, z).
?ancestor(j, r).
\end{verbatim}
\[\Phi_{family} = [+parent(d, j)]+[+parent(z, d)]+[+parent(s, z)]+[+parent(n, s)]+\]
\[[+ancestor(x, y), -parent(x, y)]+[+ancestor(x, z), -parent(x, y), -ancestor(y, z)]+\]
\[[-ancestor(j, r), r]\]
\end{example}

\subsection{Evaluation of constellations}

We are now interested in tiling evaluation, which is not standard in the theory of tiling models but lead in logic programming in our case.

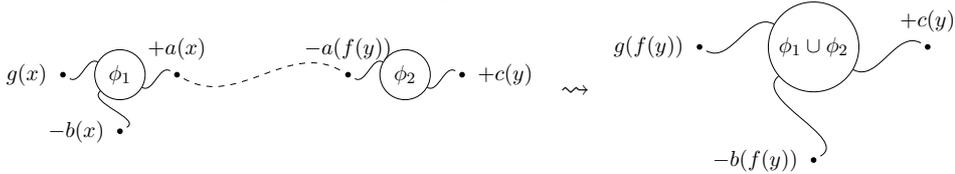
\begin{minipage}{0.5\textwidth}
\scalebox{0.75}{
\begin{tikzpicture}
    \node[circle,draw=black] at (0, 0) (c1) {$\phi_1$};
    \node[dot, label={180:$g(x)$}] at (-1, 0) (b1) {};
    \node[dot, label={90:$+a(x)$}] at (1, 0) (b2) {};
    \node[dot, label={180:$-b(x)$}] at (0, -1) (b3) {};
    \draw (c1) to[in=-30, out=150] (b1);
    \draw (c1) to[out=-30, in=150] (b2);
    \draw (c1) to[out=220, in=30] (b3);

    \node[circle,draw=black] at (5, 0) (c2) {$\phi_2$};
    \node[dot, label={90:$-a(f(y))$}] at (4, 0) (a1) {};
    \node[dot, label={0:$+c(y)$}] at (6, 0) (a2) {};
    \draw (c2) to[in=-30, out=150] (a1);
    \draw (c2) to[out=-30, in=150] (a2);

    \draw[dashed] (b2) to[out=-30, in=150] (a1);
\end{tikzpicture}}
\end{minipage}
${\Huge\leadsto}$
\begin{minipage}{0.4\textwidth}
\scalebox{0.75}{
\begin{tikzpicture}
    \node[circle,draw=black] at (0, 0) (c1) {$\phi_1 \cup \phi_2$};
    \node[dot, label={180:$g(f(y))$}] at (-2, 0) (b1) {};
    \node[dot, label={90:$+c(y)$}] at (2, 0) (b2) {};
    \node[dot, label={180:$-b(f(y))$}] at (0, -2) (b3) {};
    \draw (c1) to[in=-30, out=150] (b1);
    \draw (c1) to[out=-30, in=150] (b2);
    \draw (c1) to[out=220, in=30] (b3);
\end{tikzpicture}}
\end{minipage}

\medskip
We first define an evaluation of diagrams called \emph{actualisation}. If stars are understood as molecules, then evaluating diagrams corresponds to triggering the actual interaction between the stars along their connected rays, thus making a sort of chemical reaction happen and propagate to the non-involved entities. Another way to see it is to solve constraints between the connected rays and propagating the solution to the free (unconnected) rays (which survive to the evaluation).

There are two equivalent ways of contracting diagrams into a star by observing that each edge defines a unification problem (equation between terms):
\begin{description}
    \item[Fusion] We can reduce the links step-by-step by solving the underlying equation, producing a solution $\theta$. The two linked stars will merge by making the connected rays disappear. The substitution $\theta$ is finally applied on the rays of the resulting star. This is Robinson's resolution rule.
    \item[Actualisation] The set of all edges defines a big unification problem. The solution $\theta$ of this problem is then applied on the star of free rays.
\end{description}

In order to get a successful evaluation we need our diagrams to satisfy few properties (only the first and last ones are mandatory). We usually require diagrams to be:
\begin{description}
    \item[Connected] because otherwise, we would always have infinitely many diagrams by repeating isolated stars (notice that it ensures a reduction into a single star);
    \item[Saturated] meaning that it is impossible to extend the diagram with more occurrences of stars. It represents the idea of maximal/complete computation. In practice, we often excludes of diagrams with free coloured rays since they represent incomplete computation (see subsection \ref{subsec:automata}) but keep the definition more general when this features is not needed.
    \item[Correct] meaning that the actualisation does not fail (no contradiction during conflict resolution -- such as in the equation $f(t) \eqq g(t)$).
\end{description}

\begin{example}
\label{ex:diagrams}
We illustrate diagrams for the unary addition. Notice that diagrams correspond to traces of programs where the re-use of a star corresponds to a loop in programming.

\bigskip
Partial computation of $2+2$ (0 recursion):

\medskip
\begin{tikzpicture}[every node/.style={scale=0.8}]
    \tstar{s1}{(0, 0)}{{{$-add(0,y,y)$}}}
    \tstar{s2}{(3, 0)}{{{$+add(x,y,z)$}, {$-add(s(x),y,s(z))$}}}
    \tstar{s3}{(9, 0)}{{{$+add(s^2(0),s^2(0),r)$}, {$r$}}}
    \draw (s1/0) -- (s2/0);
    \draw (s2/1) -- (s3/0);
\end{tikzpicture}

\bigskip
Complete computation of $2+2$ (1 recursion):

\medskip
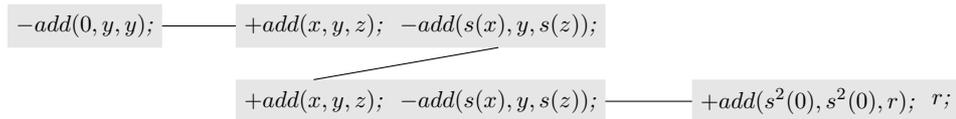
\begin{tikzpicture}[every node/.style={scale=0.8}]
    \tstar{s1}{(0, 0)}{{{$-add(0,y,y)$}}}
    \tstar{s2}{(3, 0)}{{{$+add(x,y,z)$}, {$-add(s(x),y,s(z))$}}}
    \tstar{cs2}{(3, -1)}{{{$+add(x,y,z)$}, {$-add(s(x),y,s(z))$}}}
    \tstar{s3}{(9, -1)}{{{$+add(s^2(0),s^2(0),r)$}, {$r$}}}
    \draw (s1/0) -- (s2/0);
    \draw (s2/1.south) -- (cs2/0.north);
    \draw (cs2/1) -- (s3/0);
\end{tikzpicture}

\bigskip
Over computation of $2+2$:

\medskip
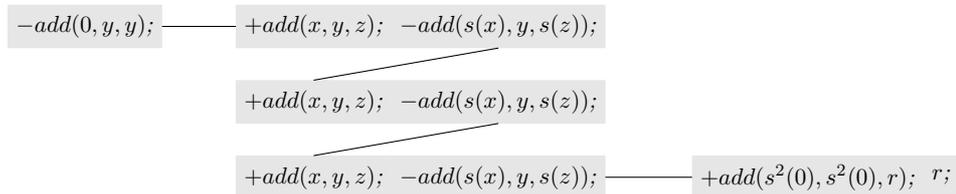
\begin{tikzpicture}[every node/.style={scale=0.8}]
    \tstar{s1}{(0, 0)}{{{$-add(0,y,y)$}}}
    \tstar{s2}{(3, 0)}{{{$+add(x,y,z)$}, {$-add(s(x),y,s(z))$}}}
    \tstar{cs2}{(3, -1)}{{{$+add(x,y,z)$}, {$-add(s(x),y,s(z))$}}}
    \tstar{ccs2}{(3, -2)}{{{$+add(x,y,z)$}, {$-add(s(x),y,s(z))$}}}
    \tstar{s3}{(9, -2)}{{{$+add(s^2(0),s^2(0),r)$}, {$r$}}}
    \draw (s1/0) -- (s2/0);
    \draw (s2/1.south) -- (cs2/0.north);
    \draw (cs2/1.south) -- (ccs2/0.north);
    \draw (ccs2/1) -- (s3/0);
\end{tikzpicture}

\bigskip
Note that in the case of $\Phi_\nat^{n,m}$, there is infinitely many saturated diagrams but only one is correct: the one corresponding to a successful computation of $n+m$.
\end{example}

\begin{example}[Fusion]
The full fusion of the diagram representing a complete computation of $2+2$ from example \ref{ex:diagrams} is described below (we make the exclusion of variable explicit):

\begin{center}
\begin{tikzpicture}[every node/.style={scale=0.8}]
    \tstar{s1}{(0, 0)}{{{$-add(0,y_1,y_1)$}}}
    \tstar{s2}{(3, 0)}{{{$+add(x_2,y_2,z_2)$}, {$-add(s(x_2),y_2,s(z_2))$}}}
    \tstar{cs2}{(3, -1)}{{{$+add(x_3,y_3,z_3)$}, {$-add(s(x_3),y_3,s(z_3))$}}}
    \tstar{s3}{(9, -1)}{{{$+add(s^2(0),s^2(0),r)$}, {$r$}}}
    \draw (s1/0) -- (s2/0);
    \draw[Red] (s2/1.south) -- (cs2/0.north);
    \draw (cs2/1) -- (s3/0);
\end{tikzpicture}
\end{center}

\begin{center}$\downarrow \theta = \{x_3 \mapsto s(x_2), y_3 \mapsto y_2, z_3 \mapsto s(z_2)\}$\end{center}

\begin{center}
\begin{tikzpicture}[every node/.style={scale=0.8}]
    \tstar{s1}{(0, 0)}{{{$-add(0,y_1,y_1)$}}}
    \tstar{s2}{(3, 0)}{{{$+add(x_2,y_2,z_2)$}, {$-add(s(s(x_2)),y_2,s(s(z_2)))$}}}
    \tstar{s3}{(10, 0)}{{{$+add(s^2(0),s^2(0),r)$}, {$r$}}}
    \draw[Red] (s1/0) -- (s2/0);
    \draw (s2/1) -- (s3/0);
\end{tikzpicture}
\end{center}

\begin{center}$\downarrow \theta = \{x_2 \mapsto 0, z_2 \mapsto y_2\}$\end{center}

\begin{center}
\begin{tikzpicture}[every node/.style={scale=0.8}]
    \tstar{s2}{(3, 0)}{{{$-add(s(s(0)),y_2,s(s(y_2)))$}}}
    \tstar{s3}{(9, 0)}{{{$+add(s^2(0),s^2(0),r)$}, {$r$}}}
    \draw[Red] (s2/0) -- (s3/0);
\end{tikzpicture}
\end{center}

\begin{center}$\downarrow \theta = \{y_2 \mapsto s(s(0)), r \mapsto s(s(s(s(0))))\}$\end{center}

\begin{center}
\begin{tikzpicture}[every node/.style={scale=0.8}]
    \tstar{s3}{(9, 0)}{{$s(s(s(s(0))))$}}
\end{tikzpicture}
\end{center}

\end{example}

\begin{example}[Actualisation]
If we take the diagram $\delta$ representing a complete computation in the example \ref{ex:diagrams}, it generates the following unification problem:
\[\mathcal{P}(\delta) = \{add(0, y_1, y_1) \eqq add(x_2, y_2, z_2), add(s(x_2), y_2, s(z_2)) \eqq add(x_3, y_3, z_3),\]
\[add(s(x_3), y_3, s(z_3)) \eqq add(s^2(0), s^2(0), r)\}\]
which is solved by a unification algorithm such as the Montanari-Martelli algorithm \cite{unifalgo}:
\[\rightarrow^* \{x_2 \eqq 0, y_2 \eqq y_1, z_2 \eqq y_1, x_3 \eqq s(x_2), y_2 \eqq y_3, z_3 \eqq s(z_2),\]
\[s(x_3) \eqq s^2(0), y_2 \eqq s^2(0), s(z_3) \eqq r\}\]
\[\rightarrow^* \{y_2 \eqq y_1, z_2 \eqq y_1, x_3 \eqq s(0), y_2 \eqq y_3, z_3 \eqq s(z_2),s(x_3) \eqq s^2(0), y_2 \eqq s^2(0), s(z_3) \eqq r\}\]
\[\rightarrow^* \{z_2 \eqq y_1, x_3 \eqq s(0), y_1 \eqq y_3, z_3 \eqq s(z_2),s(x_3) \eqq s^2(0), y_2 \eqq s^2(0), s(z_3) \eqq r\}\]
\[\rightarrow^* \{x_3 \eqq s(0), y_1 \eqq y_3, z_3 \eqq s(y_1),s(x_3) \eqq s^2(0), y_1 \eqq s^2(0), s(z_3) \eqq r\}\]
\[\rightarrow^* \{y_1 \eqq y_3, z_3 \eqq s(y_1),s(s(0)) \eqq s^2(0), y_1 \eqq s^2(0), s(z_3) \eqq r\}\]
\[\rightarrow^* \{z_3 \eqq s(y_3),s(s(0)) \eqq s^2(0), y_3 \eqq s^2(0), s(z_3) \eqq r\}\]
\[\rightarrow^* \{s(s(0)) \eqq s^2(0), y_3 \eqq s^2(0), s(s(y_3)) \eqq r\}\]
\[\rightarrow^* \{y_3 \eqq s^2(0), s(s(y_3)) \eqq r\}\]
\[\rightarrow^* \{s(s(s^2(0))) \eqq r\}\]
\[\rightarrow^* \{r \eqq s(s(s^2(0)))\}\]
The solution of this problem is the substitution $\theta = \{r \mapsto s^4(0)\}$ which is applied on the star of free rays $[r]$. The result $[s^4(0)]$ of this procedure is called the \emph{actualisation} of $\delta$.
\end{example}

\[\Phi
\quad\overset{\text{generates}}{\longrightarrow}\quad \cup^\infty_{k=1} D_k
\quad\overset{\text{actualises}}{\longrightarrow}
\phi_1 + ... + \phi_n\]

The \emph{execution} or \emph{normalisation} $\exec(\Phi)$ of a constellation $\Phi$ constructs the set of all possible correct saturated diagrams and actualises them all in order to produce a new constellation called the \emph{normal form}.

In logic programming, we can interpret the normal form as a subset of the application of resolution operator \cite{leitsch2012resolution}. It non-deterministically computes all the "maximal/complete" inferences possible.

If the set of correct saturated diagrams is finite (or the normal form is a finite constellation), the constellation is said to be \emph{strongly normalising}.

\begin{example}
For $\Phi_\nat^{2,2}$ (from Example~\ref{ex:logicprograms}), one can check that we have $\exec(\Phi_\nat^{2,2}) = [s^4(0)]$ because only the complete computation of example \ref{ex:diagrams} succeed and all other saturated diagrams representing partial or over computations fail.
\end{example}

\section{Encoding of some models of computation}

We provide concrete examples of how our model computes. There are two interesting facts about computing with stellar resolution.
\begin{itemize}
    \item The stellar resolution computes by encoding a structure (hypergraph) and a transmission of information inside it. This generalises circuits and automata. It also makes the correspondence between tile systems and automata direct (some links were already investigated in the literature \cite{thomas1991logics}).
    \item Models of computation dependent of external definitions (the evaluation function of circuits for instance) are translated into two dual constellations, one corresponding to the model and the other corresponding to its environment, semantics or external evaluation function. We require that they interact as expected. This will be illustrated in the encoding of circuits.
\end{itemize}

\subsection{Non-deterministic finite automata}
\label{subsec:automata}

The idea is to represent the transitions by binary stars with two opposite polarities (hence representing an edge in a state graph).

Let $\Sigma$ be an alphabet and $w \in  \Sigma^*$ a word. If $w = c_1...c_n$ then $w^\bigstar = [+i(c_1 \cdot (... \cdot (c_n \cdot \varepsilon)))]$. We use the binary function symbol $\cdot$ considered right-associative.

Let $A = (\Sigma, Q, Q_0, \delta, F)$ be a non-deterministic finite automata. We define its translation $A^\bigstar$:
\begin{itemize}
    \item for each $q_0 \in Q_0$, we have $[-i(w), +a(w, q_0)]$.
    \item for each $q_f \in F$, we have $[-a(\varepsilon, q_f), \mathtt{accept}]$.
    \item for each $q \in Q, c \in \Sigma$ and for each $q' \in \delta(q, c)$, we have the star
    \[[-a(c \cdot w, q), +a(w, q')].\]
\end{itemize}

For instance, the following automaton $A$ accepting binary words ending by $00$:
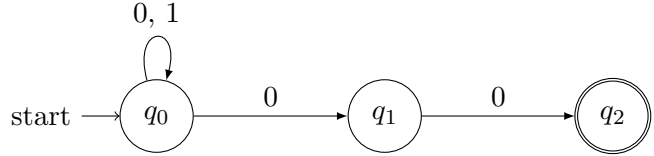
\begin{center}
    \begin{tikzpicture}
        \node[state, initial] (q0) {$q_0$};
        \node[state, right of=q0, xshift=20mm] (q1) {$q_1$};
        \node[state, accepting, right of=q1, xshift=20mm] (q2) {$q_2$};
        \draw[-latex]
        (q0) edge[loop above] node{0, 1} (q0)
        (q0) edge[above] node{0} (q1)
        (q1) edge[above] node{0} (q2);
    \end{tikzpicture}
\end{center}

is translated as:
\[A^\bigstar = [-i(w), +a(w, q_0)]+[-a(\epsilon, q_2), \mathtt{accept}]+\]
\[[-a(0 \cdot w, q_0), +a(w, q_0)]+[-a(1 \cdot w, q_0), +a(w, q_0)]+\]
\[[-a(0 \cdot w, q_0), +a(w, q_1)]+[-a(0 \cdot w, q_1), +a(w, q_2)]\]

The set of saturated correct diagrams corresponds to the set of non-deterministic runs. With the word $[+i(0 \cdot 0 \cdot 0 \cdot \varepsilon)]$ only the run $q_0q_0q_1q_2$ leads to the accepting state $q_2 \in F$. The corresponding diagram will actualise into $[\mathtt{accept}]$. The other correct diagrams correspond to incomplete computations and are excluded. We get $\lightning\exec(A^\bigstar+[+i(0 \cdot 0 \cdot 0 \cdot \varepsilon)]) = [\mathtt{accept}]$ where $\lightning\Phi$ erases the stars containing coloured rays (so that we delete incomplete garbage paths from the result), meaning that the word is accepted.

This idea can easily be extended to pushdown automata. We briefly illustrate the idea: the star $[-a(1 \cdot w, 0 \cdot s), +a(w, s)]$ corresponds to checking if we read $1$ and that $0$ is on the top of the stack and if so, we remove it. Since we manipulate terms and that Turing machines can be seen as pushdown automata extended with two stacks, the extension to tree automata and Turing machines is direct.

Also remark that there is no explicit flow of computation and that paths are constructed non-deterministically in an asynchronous and concurrent way.

\subsection{Boolean circuits}

The idea is to first encode a hypergraph representing the structure of a boolean circuit then to connect the resulting constellation to another one containing the implementation of logic gates (the semantics) as if it was a sort of \texttt{\#include <proplogic.h>} in the C language.

\[VAR(Y, i) := [-val(x), Y(x), +c_i(x)]\]
\[SHARE(i, j, k) := [-c_{i}(x), +c_{j}(x), +c_{k}(x)]\]
\[AND(i, j, k) := [-c_{i}(x), -c_{j}(y), -and(x, y, r), +c_{k}(r)]\]
\[OR(i, j, k) := [-c_{i}(x), -c_{j}(y), -or(x, y, r), +c_{k}(r)]\]
\[NEG(i, j) := [-c_{i}(x), -neg(x, r), +c_{j}(r)]\]
\[CONST(k, i) := [+c_i(k)] \qquad QUERY(k, i) := [+c_i(k), R(k)]\]
where $i, j, k$ are encodings of natural numbers representing identifiers. We also have a star $VAR(Y, i)$ for each variable $Y$ we want as input in our boolean circuit.

We consider the following constellation representing a "module" (as in any programming language) providing the definitions of propositional logic:
\[\Phi^{\mathcal{PL}} = [+val(0)] + [+val(1)] +
[+neg(0, 1)] + [+neg(1, 0)] +
\]
\[
[+and(0, 0, 0)] + [+and(0, 1, 0)] + [+and(1, 0, 0)] + [+and(1, 1, 1)]
\]
\[
[+or(0, 0, 0)] + [+or(0, 1, 1)] + [+or(1, 0, 1)] + [+or(1, 1, 1)].
\]

We can observe that by changing the module $\Phi^{\mathcal{PL}}$, we can adapt the encoding and obtain arithmetic circuits (possibly with several alternative implementations) or even consider several implementations of logic/arithmetic gates (for instance with $[+or(0,x,x)]+[+or(1,x,1)]$ for the OR gate). Notice that syntax and semantics live in the same world (they are represented as objects of the same kind) and can interact.

We use the star $CONST(k)$ to force a value on the input and the star $QUERY(k)$ to ask for a particular output. For instance, $QUERY(1)$ asks for satisfiability.

To illustrate the encoding, we execute the constellation representing $X \lor \lnot X$ where $\underline{n}$ is an encoding of natural number:
\[\Phi_{em} = VAR(X,\underline{0})+SHARE(\underline{0},\underline{1},\underline{2})+NEG(\underline{2},\underline{3})+OR(\underline{1},\underline{3},\underline{4})+QUERY(1, \underline{4})\]

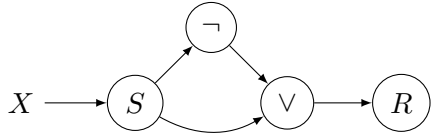
\begin{center}
    \begin{tikzpicture}
        \node at (0, 0) (x) {$X$};
        \node[circle, draw] at (1.5, 0) (s) {$S$};
        \node[circle, draw] at (2.5, 1) (not) {$\lnot$};
        \node[circle, draw] at (3.5, 0) (or) {$\lor$};
        \node[circle, draw] at (5, 0) (r) {$R$};
        \draw[-latex] (x) -- (s);
        \draw[-latex] (s) to (not);
        \draw[-latex] (s) to[bend right] (or);
        \draw[-latex] (not) -- (or);
        \draw[-latex] (or) -- (r);
    \end{tikzpicture}
\end{center}

The constellation $\Phi_{em} \uplus \Phi^{\mathcal{PL}}$ will generate two diagrams: one corresponding to the input $0$ and another one for $1$. We obtain $\exec(\Phi_{em}+\Phi^{\mathcal{PL}}) = [X(0), R(1)]+[X(1), R(1)]$ stating that for the two valuations $x\mapsto 0$ and $x\mapsto 1$, the circuit outputs $R(1)$.

\subsection{Comments on the model}

This model of computation embodies the idea of computation-as-interaction \cite{abramsky2008information}. It is a very primitive model of computation which is based on our intuition of space (a geometric/topological structure where information flows). The rays can be seen as wires we can peel (for instance $f(x)$ can be divided into subwires $f(\mathtt{l} \cdot x)$ and $f(\mathtt{r} \cdot x)$) and thus opening topological/geometrical interpretations.

\begin{description}
    \item[Relationship with logic programming] At first glance, our model is identical to Robinson's first-order resolution using disjunctive clauses. The difference is that our model is purely computational (no reference to logic) and that we use it for a different purpose (no interest in reaching the empty clause but rather the set of atoms we can infer). We can also have information not involved in computation (unpolarised rays) which collects data which will be outputted in the normal form. Although this dynamics of logic programming is well-known and well-understood it seems that it has never been used that way as a logic-free concurrent constraint programming language\footnote{I believe that it is simply because no practical purpose would justify such a model of computation.}. Moreover, our model will be extended with internal colours and possibly additional features in the future, thus justifying the use of a new name.
    \item[Relationship with tile systems] The stellar resolution generalises \emph{flexible tiles} \cite{jonoska2006flexible, jonoska2005computational}, a model of computation coming from DNA computing \cite{winfree1998algorithmic}. This model is itself able to simulate tile systems such as Wang tiles \cite{wang} or abstract tile assembly models \cite{patitz2014introduction} by encoding both the tile set and its environment by constellations. From our interactive typing, one can imagine methods of typing and implicit complexity analysis of these models.
    \item[Relationship with the GoI and complexity] We can generalise flows \cite{goi3, bagnol2014resolution} and interaction graphs \cite{seillerPHD, seiller2016interaction2} which have applications in implicit computational complexity \cite{baillot2001elementary, aubert2014unification, aubert2016unary, seiller2018interaction}. Some of these works encode some notions of automata which we can also reproduce and extend.
\end{description}

\section{Geometry of Interaction for MLL}

We explain how to encode MLL proof-structures and how to simulate both MLL cut-elimination and logical correctness (by the Danos-Regnier criterion) as a first step towards a full reconstruction of linear logic.

\subsection{Cut-elimination and permutations}

In a proof-structure, axioms induce a permutation on distinct natural numbers representing atoms (the conclusion of axiom rules). These atoms (called \emph{loci}, which is the plural of \emph{locus}, in ludics \cite{girard2001locus}) represent the physical locations involved in a proof (you can think of computer memory). We can actually forget the formulas which are simply syntactic labels and only retain these locations without any special logical meaning.

The $\parr/\otimes$ cuts can be seen as administrative/inessential cuts since all they do is basically a rewiring on the premises of the $\parr$ and $\otimes$ nodes connected together. Actually, when considering expansed axioms, they can all be eliminated in order to produce a canonical form with only cut between atoms. Therefore, cuts can be seen as a (partial) permutation on atoms. The atoms become the ``support of interaction'' of the proof; the locations where logical interaction occurs. Here is an illustration of two permutations representing a proof-structure with cuts and atoms in $\{1, 2, 3, 4, 5, 6, 7, 8\}$:

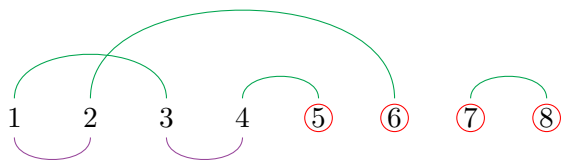
\begin{center}
    \begin{tikzpicture}
    \foreach \i in {1,...,8}
    {
        \node at (\i-1, 0) (n\i) {$\i$};
    }
    \node[circle, draw, red] at (n5.center) {};
    \node[circle, draw, red] at (n6.center) {};
    \node[circle, draw, red] at (n7.center) {};
    \node[circle, draw, red] at (n8.center) {};

    \draw[Green] (n1.north) edge[bend left, in=90, out=90] (n3.north);
    \draw[Green] (n2.north) edge[bend left, in=90, out=90] (n6.north);
    \draw[Green] (n7.north) edge[bend left, in=90, out=90] (n8.north);
    \draw[Green] (n4.north) edge[bend left, in=90, out=90] (n5.north);

    \draw[Purple] (n1.south) edge[bend right, in=-90, out=-90] (n2.south);
    \draw[Purple] (n3.south) edge[bend right, in=-90, out=-90] (n4.south);
    \end{tikzpicture}
\end{center}

A simple presentation of GoI is given by Seiller \cite{seillerPHD} with connexion between graphs instead of permutations on a set of natural numbers representing atoms. These graphs are called \emph{interaction graphs} for their ability to connect with each other. The cut-elimination becomes the computation of maximal alternating path between two graphs (the graph of axioms and the graph of cuts). This is actually the same as considering an edge contraction as in the cut-elimination of proof-structures. In this new framework, proof-nets theory is about locations and paths, and truth or logical correctness is about cycles and connectivity, hence purely structural considerations.

The stellar resolution generalises this idea by encoding hypergraphs representing proof-structures. In the case of cut-elimination, we only need binary stars representing graph edges. The above interaction graph becomes:

\[\Phi := [\ccol{+c(1)}, \ccol{-c(3)}] + [\ccol{+c(4)}, 5] + [\ccol{+c(2)}, 6] + [7, 8]\]
\[[\ccol{-c(1)}, \ccol{-c(2)}] + [\ccol{-c(3)}, \ccol{-c(4)}]\]

We have $\exec(\Phi) = [5, 6]+[7, 8]$ which corresponds to the expected normal form (set of maximal paths). Notice that the matching is exact, hence the actualisation is a trivial contraction of a unique diagram (no non-deterministic choice is involved).

We will see later a more sophisticated version which the reproduce the behaviour of $\otimes/\parr$ cuts so that we do not necessarily have to consider normalised proof-structures.

\subsection{Correctness and partitions}

If we wish to treat logical correctness in a satisfactory and natural way for MLL, we have to shift to partitions of a set instead of permutations \cite{girard1987multiplicatives, acclavio2020generalized}. Permutations can still be retrieved: a permutation $\{x_1 \mapsto y_1, ..., x_n \mapsto y_n\}$ on $X \subseteq \nat$ representing a proof naturally induces a partition $\{\{x_1, y_1\}, ..., \{x_n, y_n\}\}$ in $X$.

A Danos-Regnier switching induces a partition depending on how it separates or groups atoms into different connected components:

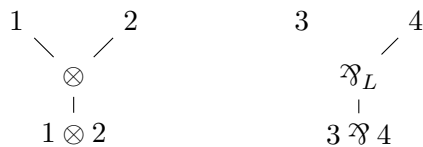
\begin{center}
\begin{tikzpicture}
  % tensor
  \node at (-0.75, 0.75) (a) {$1$};
  \node at (0.75, 0.75) (b) {$2$};
  \node at (0, 0) (tens) {$\otimes$};
  \node at (0, -0.75) (ab) {$1 \otimes 2$};
  \draw[-] (a) -- (tens);
  \draw[-] (b) -- (tens);
  \draw[-] (tens) -- (ab);

  % par
  \node at (3, 0.75) (ad) {$3$};
  \node at (4.5, 0.75) (bd) {$4$};
  \node at (3.75, 0) (par) {$\parr_L$};
  \node at (3.75, -0.75) (adbd) {$3 \parr 4$};
  \draw[-] (bd) -- (par);
  \draw[-] (par) -- (adbd);
\end{tikzpicture}
\end{center}

The above switching graph corresponds to the partition $\{\{1, 2\}, \{3\}, \{4\}\}$ because the tensor groups the atoms $1$ and $2$ together while $\parr_L$ separates $3$ and $4$. Partitions are related by \emph{orthogonality}: two partitions are orthogonal when the graph constructed with sets as nodes and where two nodes are adjacent whenever they share a common value, is a tree. Testing a partition coming from axioms against several partitions for all the switching graphs is sufficient to speak about logical correctness. For more details, see Acclavio and Maieli's works \cite{acclavio2020generalized}.

Since stars are not limited to the binary case, we can naturally represent general partitions by constellations. The above switching graph is translated into the following constellation: $[\ccol{-c(1)}, \ccol{-c(2)}]+[\ccol{-c(3)}]+[\ccol{-c(4)}]$ (notice the negative polarity in order to allow connexion with axioms). However, in this case, all diagrams are closed (no free rays). For technical reasons, because otherwise the switchings would not be distinguished, we have to specify where the conclusions are located: $[\ccol{-c(1)}, \ccol{-c(2)}, 1 \otimes 2]+[\ccol{-c(3)}]+[\ccol{-c(4)}, 3 \parr 4]$.

We finally obtain the two following constellations $\Phi$ for axioms and $\Phi_\parr^L$ for the above switching graph:
\[\Phi = [\ccol{+c(1)}, \ccol{+c(3)}]+[\ccol{+c(2)}, \ccol{+c(4)}]\]
\[\Phi_\parr^L = [\ccol{-c(1)}, \ccol{-c(2)}, 1 \otimes 2]+[\ccol{-c(3)}]+[\ccol{-c(4)}, 3 \parr 4].\]

We have $\exec(\Phi \uplus \Phi_\parr^L) = [1 \otimes 2, 3 \parr 4]$ which is the star of conclusions. In this case, we say that $\Phi$ passes the test $\Phi_\parr^L$.

The Danos-Regnier criterion is reformulated as follows: \emph{``a constellation representing a proof-structure is correct if and only if its execution against all the constellations representing its switching graphs produces the star of its conclusions"}.

\section{Stellar interpretation of multiplicative linear logic}

\subsection{The computational content of multiplicatives}

We now consider a more general setting using more sophisticated terms as rays instead of simple constants. This is not especially useful for MLL but will be crucial for the exponentials or further extensions (for instance first and second-order linear logic).

We set a basis of representation $\mathbb{B}$ with unary symbols $p_A$ for all formulas $A$ of MLL, and constants $\mathtt{l}, \mathtt{r}$ used to encode the addresses of an atom relatively to the arborescence of the lower part of a proof-structure. We use a right-associative binary symbol $\cdot$ to glue constants together. Any other isomorphic basis can be considered as well.

To simulate the dynamics of cut-elimination we translate the axioms and the cuts into stars:

\begin{enumerate}
    \item An atom $A$ becomes a ray $\ccol{+c.p_C(t)}$ where $t$ represents the "address" of $A$ relatively to a conclusion $C$ of the proof-structure (without considering cuts). We use and encoding of the path from a conclusion to $A$ in the tree corresponding to the lower part of the proof-structure. For instance, for a path going twice left from a conclusion $C$, we have the address $t := \mathtt{l} \cdot \mathtt{l} \cdot x$. The colour $\ccol{+c}$ stands for "positive computation".
    \item An axiom becomes a binary star containing the translations of its atoms as described above. For instance, the axiom $\vdash A, A^\bot$ becomes $[\ccol{+c.p_A(x)}, \ccol{+c.p_{A^\bot}(x)}]$.
    \item A cut between $A$ and $A^\bot$ becomes a binary star $[\ccol{-c.p_A(x)}, \ccol{-c.p_B(x)}]$ coloured with $\ccol{-c}$ for "negative computation".
\end{enumerate}

\begin{example}
We encode the following cut-elimination $\mathcal{S} \rightarrow^* \mathcal{S}'$ of MLL proof-structures:

\medskip
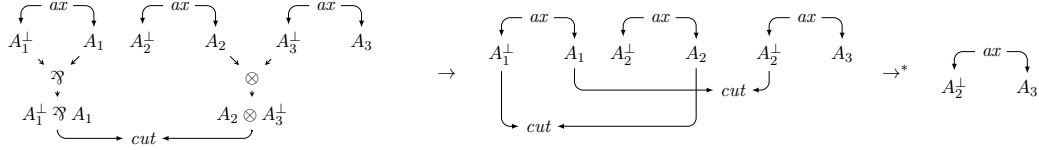
\scalebox{0.64}{
\begin{minipage}{0.6\textwidth}
\begin{tikzpicture}
    \node at (-0.75, 0.75) (a) {$A_1^\bot$};
    \node at (0.75, 0.75) (b) {$A_1$};
    \node at (0, 0) (par) {$\parr$};
    \node at (0, -0.75) (ab) {$A_1^\bot \parr A_1$};
    \draw[-stealth] (a) -- (par);
    \draw[-stealth] (b) -- (par);
    \draw[-stealth] (par) -- (ab);

    \node at (1.75, 0.75) (f1) {$A_2^\bot$};
    \node at (6.25, 0.75) (f2) {$A_3$};

    \node at (3.25, 0.75) (c) {$A_2$};
    \node at (4.75, 0.75) (d) {$A_3^\bot$};
    \node at (4, 0) (tens) {$\otimes$};
    \node at (4, -0.75) (cd) {$A_2 \otimes A_3^\bot$};
    \draw[-stealth] (c) -- (tens);
    \draw[-stealth] (d) -- (tens);
    \draw[-stealth] (tens) -- (cd);

    \node at (1.75, -1.25) (cut) {cut};
    \draw[-latex, rounded corners=5pt] (ab) |- (cut);
    \draw[-latex, rounded corners=5pt] (cd) |- (cut);
    \node at (0, 1.5) (ax) {ax};
    \draw[-latex, rounded corners=5pt] (ax) -| (a);
    \draw[-latex, rounded corners=5pt] (ax) -| (b);
    \node at (2.5, 1.5) (ax) {ax};
    \draw[-latex, rounded corners=5pt] (ax) -| (f1);
    \draw[-latex, rounded corners=5pt] (ax) -| (c);
    \node at (5.5, 1.5) (ax) {ax};
    \draw[-latex, rounded corners=5pt] (ax) -| (f2);
    \draw[-latex, rounded corners=5pt] (ax) -| (d);
\end{tikzpicture}
\end{minipage}
$\rightarrow\quad$
\begin{minipage}{0.55\textwidth}
\begin{tikzpicture}
    \node at (-0.75, 0.75) (a) {$A_1^\bot$};
    \node at (0.75, 0.75) (b) {$A_1$};

    \node at (1.75, 0.75) (f1) {$A_2^\bot$};
    \node at (6.25, 0.75) (f2) {$A_3$};

    \node at (3.25, 0.75) (c) {$A_2$};
    \node at (4.75, 0.75) (d) {$A_2^\bot$};

    \node at (0, -0.75) (cut) {cut};
    \draw[-latex, rounded corners=5pt] (a) |- (cut);
    \draw[-latex, rounded corners=5pt] (c) |- (cut);
    \node at (4, 0) (cut) {cut};
    \draw[-latex, rounded corners=5pt] (b) |- (cut);
    \draw[-latex, rounded corners=5pt] (d) |- (cut);
    \node at (0, 1.5) (ax) {ax};
    \draw[-latex, rounded corners=5pt] (ax) -| (a);
    \draw[-latex, rounded corners=5pt] (ax) -| (b);
    \node at (2.5, 1.5) (ax) {ax};
    \draw[-latex, rounded corners=5pt] (ax) -| (f1);
    \draw[-latex, rounded corners=5pt] (ax) -| (c);
    \node at (5.5, 1.5) (ax) {ax};
    \draw[-latex, rounded corners=5pt] (ax) -| (f2);
    \draw[-latex, rounded corners=5pt] (ax) -| (d);
\end{tikzpicture}
\end{minipage}
$\rightarrow^*\quad$
\begin{minipage}{0.1\textwidth}
\begin{tikzpicture}
    \node at (-0.75, 0.75) (a) {$A_2^\bot$};
    \node at (0.75, 0.75) (b) {$A_3$};
    \node at (0, 1.5) (ax) {ax};
    \draw[-latex, rounded corners=5pt] (ax) -| (a);
    \draw[-latex, rounded corners=5pt] (ax) -| (b);
\end{tikzpicture}
\end{minipage}}

\medskip
The address of $A_1^\bot$ is $p_{A_1^\bot \parr A_1}(\mathtt{l} \cdot x)$ because it is located on the left-hand side of $A_1^\bot \parr A_1$. The address of $A_3^\bot$ is $p_{A_2 \otimes A_3^\bot}(\mathtt{r} \cdot x)$ and the one for $A_3$ is $p_{A_3}(x)$. The proof-structure $\mathcal{S}$ is encoded as:
\[
[\ccol{+c.p_{A_1^\bot \parr A_1}(\mathtt{l} \cdot x)}, \ccol{+c.p_{A_1^\bot \parr A_1}(\mathtt{r} \cdot x)}]+
[\ccol{+c.p_{A_2^\bot}(x)}, \ccol{+c.p_{A_2 \otimes A_3^\bot}(\mathtt{l} \cdot x)}]+
\]
\[
[\ccol{+c.p_{A_2 \otimes A_3^\bot}(\mathtt{r} \cdot x)}, \ccol{+c.p_{A_3^\bot}(x)}]+
[\ccol{-c.p_{A_1^\bot \parr A_1}(x)}, \ccol{-c.p_{A_2 \otimes A_3^\bot}(x)}]
\]

The only correct saturated diagram is:

\bigskip
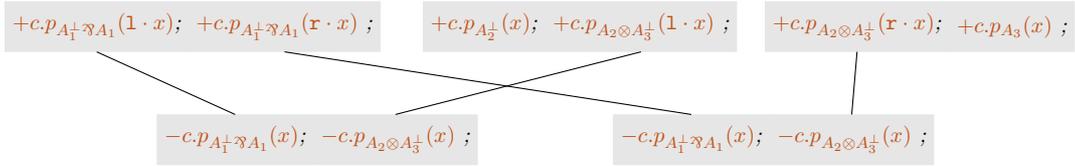
\begin{tikzpicture}[every node/.style={scale=0.75}]
    \tstar[25]{s1}{(0, 0)}{{
        {$\ccol{+c.p_{A_1^\bot \parr A_1}(\mathtt{l} \cdot x)}$},
        {$\ccol{+c.p_{A_1^\bot \parr A_1}(\mathtt{r} \cdot x)}$}
    }}
    \tstar[25]{s2}{(5.5, 0)}{{
        {$\ccol{+c.p_{A_2^\bot}(x)}$},
        {$\ccol{+c.p_{A_2 \otimes A_3^\bot}(\mathtt{l} \cdot x)}$}
    }}
    \tstar[25]{s3}{(10, 0)}{{
        {$\ccol{+c.p_{A_2 \otimes A_3^\bot}(\mathtt{r} \cdot x)}$},
        {$\ccol{+c.p_{A_3}(x)}$}
    }}
    \tstar[25]{c1}{(2, -1.5)}{{
        {$\ccol{-c.p_{A_1^\bot \parr A_1}(x)}$},
        {$\ccol{-c.p_{A_2 \otimes A_3^\bot}(x)}$}
    }}
    \tstar[25]{c2}{(8, -1.5)}{{
        {$\ccol{-c.p_{A_1^\bot \parr A_1}(x)}$},
        {$\ccol{-c.p_{A_2 \otimes A_3^\bot}(x)}$}
    }}
    \draw (s1/0.south) -- (c1/0.north);
    \draw (s1/1.south) -- (c2/0.north);
    \draw (s2/1.south) -- (c1/1.north);
    \draw (s3/0.south) -- (c2/1.north);
\end{tikzpicture}

The matching is exact (since no non-deterministic choice is involved) and when executing the constellation, we end up with $[\ccol{+c.p_{A_2^\bot}(x)}, \ccol{+c.p_{A_3}(x)}]$ corresponding to $\mathcal{S}'$, as expected.
\end{example}

\begin{example}
If we have the following reduction $\mathcal{S} \rightarrow \mathcal{S}'$ instead:

\medskip
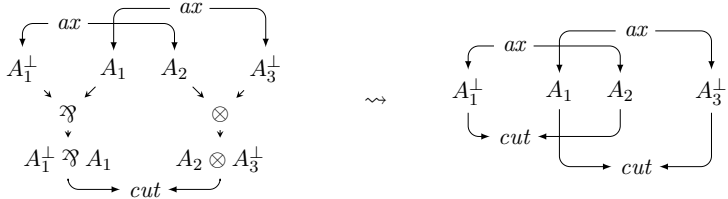
\scalebox{0.8}{
\begin{minipage}{0.4\textwidth}
\begin{tikzpicture}
    \node at (-0.75, 0.75) (a) {$A_1^\bot$};
    \node at (0.75, 0.75) (b) {$A_1$};
    \node at (0, 0) (par) {$\parr$};
    \node at (0, -0.75) (ab) {$A_1^\bot \parr A_1$};
    \draw[-stealth] (a) -- (par);
    \draw[-stealth] (b) -- (par);
    \draw[-stealth] (par) -- (ab);

    \node at (1.75, 0.75) (c) {$A_2$};
    \node at (3.25, 0.75) (d) {$A_3^\bot$};
    \node at (2.5, 0) (tens) {$\otimes$};
    \node at (2.5, -0.75) (cd) {$A_2 \otimes A_3^\bot$};
    \draw[-stealth] (c) -- (tens);
    \draw[-stealth] (d) -- (tens);
    \draw[-stealth] (tens) -- (cd);

    \node at (1.25, -1.25) (cut) {cut};
    \draw[-latex, rounded corners=5pt] (ab) |- (cut);
    \draw[-latex, rounded corners=5pt] (cd) |- (cut);
    \node at (0, 1.5) (ax) {ax};
    \draw[-latex, rounded corners=5pt] (ax) -| (a);
    \draw[-latex, rounded corners=5pt] (ax) -| (c);
    \node at (2, 1.75) (ax) {ax};
    \draw[-latex, rounded corners=5pt] (ax) -| (b);
    \draw[-latex, rounded corners=5pt] (ax) -| (d);
\end{tikzpicture}
\end{minipage}
$\leadsto\qquad$
\begin{minipage}{0.25\textwidth}
\begin{tikzpicture}
    \node at (-0.75, 0.75) (a) {$A_1^\bot$};
    \node at (0.75, 0.75) (b) {$A_1$};

    \node at (1.75, 0.75) (c) {$A_2$};
    \node at (3.25, 0.75) (d) {$A_3^\bot$};

    \node at (0, 0) (cut) {cut};
    \draw[-latex, rounded corners=5pt] (a) |- (cut);
    \draw[-latex, rounded corners=5pt] (c) |- (cut);
    \node at (2, -0.5) (cut) {cut};
    \draw[-latex, rounded corners=5pt] (b) |- (cut);
    \draw[-latex, rounded corners=5pt] (d) |- (cut);
    \node at (0, 1.5) (ax) {ax};
    \draw[-latex, rounded corners=5pt] (ax) -| (a);
    \draw[-latex, rounded corners=5pt] (ax) -| (c);
    \node at (2, 1.75) (ax) {ax};
    \draw[-latex, rounded corners=5pt] (ax) -| (b);
    \draw[-latex, rounded corners=5pt] (ax) -| (d);
\end{tikzpicture}
\end{minipage}}

\medskip
The constellation corresponding to $\mathcal{S}$ is \[
[\clocus{A_1^\bot \parr A_1}{\mathtt{l} \cdot x},
\clocus{A_2 \otimes A_3^\bot}{\mathtt{l} \cdot x}+[\clocus{A_2 \otimes A_2^\bot}{\mathtt{r} \cdot x},
\clocus{A_1^\bot \parr A_1}{\mathtt{r} \cdot x}]+\]
     \[[\vcut{A_1^\bot \parr A_1}{x}, \vcut{A_2 \otimes A_3^\bot}{x}]
\]

When trying to make a saturated diagram by following the shape of the proof-structure, we end up with:

\bigskip
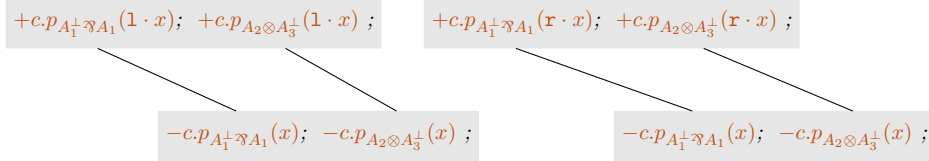
\begin{tikzpicture}[every node/.style={scale=0.75}]
    \tstar[25]{s1}{(0, 0)}{{
        {$\ccol{+c.p_{A_1^\bot \parr A_1}(\mathtt{l} \cdot x)}$},
        {$\ccol{+c.p_{A_2 \otimes A_3^\bot}(\mathtt{l} \cdot x)}$}
    }}
    \tstar[25]{s2}{(5.5, 0)}{{
        {$\ccol{+c.p_{A_1^\bot \parr A_1}(\mathtt{r} \cdot x)}$},
        {$\ccol{+c.p_{A_2 \otimes A_3^\bot}(\mathtt{r} \cdot x)}$}
    }}
    \tstar[25]{c1}{(2, -1.5)}{{
        {$\ccol{-c.p_{A_1^\bot \parr A_1}(x)}$},
        {$\ccol{-c.p_{A_2 \otimes A_3^\bot}(x)}$}
    }}
    \tstar[25]{c2}{(8, -1.5)}{{
        {$\ccol{-c.p_{A_1^\bot \parr A_1}(x)}$},
        {$\ccol{-c.p_{A_2 \otimes A_3^\bot}(x)}$}
    }}
    \draw (s1/0.south) -- (c1/0.north);
    \draw (s1/1.south) -- (c1/1.north);
    \draw (s2/0.south) -- (c2/0.north);
    \draw (s2/1.south) -- (c2/1.north);
\end{tikzpicture}

which normalises into infinitely many occurrences of the empty star $[]$. Hence, the proof-structure is incorrect.
\end{example}

\subsection{The logical content of multiplicatives}

We translate the correctness criterion of Danos-Regnier \cite{danos1989structure} by an encoding of hypergraphs (representing proof-structures) into constellations.

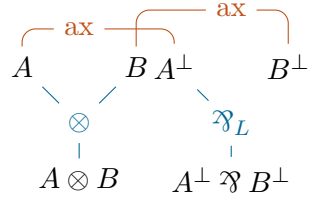
\begin{center}
\begin{tikzpicture}
    % tensor
    \node at (-0.75, 0.75) (a) {$A$};
    \node at (0.75, 0.75) (b) {$B$};
    \node[MidnightBlue] at (0, 0) (tens) {$\otimes$};
    \node at (0, -0.75) (ab) {$A \otimes B$};
    \draw[MidnightBlue, -] (a) -- (tens);
    \draw[MidnightBlue, -] (b) -- (tens);
    \draw[MidnightBlue, -] (tens) -- (ab);

    % par
    \node at (1.25, 0.75) (ad) {$A^\bot$};
    \node at (2.75, 0.75) (bd) {$B^\bot$};
    \node[MidnightBlue] at (2, 0) (par) {$\parr_L$};
    \node at (2, -0.75) (adbd) {$A^\bot \parr B^\bot$};
    \draw[MidnightBlue, -] (ad) -- (par);
    \draw[MidnightBlue, -] (par) -- (adbd);

    % ax
    \node[Bittersweet] at (0, 1.25) (ax1) {ax};
    \node[Bittersweet] at (2, 1.5) (ax2) {ax};
    \draw[Bittersweet, -, rounded corners=5pt] (ax1) -| (a);
    \draw[Bittersweet, -, rounded corners=5pt] (ax1) -| (ad);
    \draw[Bittersweet, -, rounded corners=5pt] (ax2) -| (b);
    \draw[Bittersweet, -, rounded corners=5pt] (ax2) -| (bd);
    \end{tikzpicture}
\end{center}

Any proof-structure of conclusion $\vdash A_1, ..., A_n$ can be seen as the sum of two components:
\begin{itemize}
    \item the \textcolor{Bittersweet}{upper part} made of axioms is called the \emph{vehicle};
    \item the \textcolor{MidnightBlue}{lower part} is basically the syntax forest of $\vdash A_1, ..., A_n$. The Danos-Regnier correctness criterion is obtained by considering switchings on the lower part and checking the acyclicity and connectedness when connected with axioms. This can be understood as \emph{testing} the vehicle against a \emph{set of test} (which we call \emph{format}). This testing between vehicle and format produces a certification: if all tests pass, we have a \emph{proof-net}. Note that testing is symmetric because tests are encoded as constellations as well. Moreover, a format can also be seen as tested by a vehicle.
\end{itemize}

\begin{figure}
\centering
\begin{tabular}{|c|c|c|}
    \hline
    Switching & Vehicle & Test \\
    \hline
    \scalebox{0.75}{\begin{tikzpicture}
      % tensor
      \node at (-0.75, 0.75) (a) {$A$};
      \node at (0.75, 0.75) (b) {$B$};
      \node at (0, 0) (tens) {$\otimes$};
      \node at (0, -0.75) (ab) {$A \otimes B$};
      \draw[-] (a) -- (tens);
      \draw[-] (b) -- (tens);
      \draw[-] (tens) -- (ab);

      % par
      \node at (1.25, 0.75) (ad) {$A^\bot$};
      \node at (2.75, 0.75) (bd) {$B^\bot$};
      \node at (2, 0) (par) {$\parr_L$};
      \node at (2, -0.75) (adbd) {$A^\bot \parr B^\bot$};
      \draw[-] (ad) -- (par);
      \draw[-] (par) -- (adbd);

      % ax
      \node at (0, 1.25) (ax1) {ax};
      \node at (2, 1.5) (ax2) {ax};
      \draw[-, rounded corners=5pt] (ax1) -| (a);
      \draw[-, rounded corners=5pt] (ax1) -| (ad);
      \draw[-, rounded corners=5pt] (ax2) -| (b);
      \draw[-, rounded corners=5pt] (ax2) -| (bd);
    \end{tikzpicture}}
    &
    \scalebox{0.75}{\begin{tikzpicture}
      % tensor
      \node at (-0.75, 0.75) (a) {$A$};
      \node at (0.75, 0.75) (b) {$B$};

      % par
      \node at (1.25, 0.75) (ad) {$A^\bot$};
      \node at (2.75, 0.75) (bd) {$B^\bot$};

      % ax
      \node at (0, 1.25) (ax1) {ax};
      \node at (2, 1.5) (ax2) {ax};
      \draw[-, rounded corners=5pt] (ax1) -| (a);
      \draw[-, rounded corners=5pt] (ax1) -| (ad);
      \draw[-, rounded corners=5pt] (ax2) -| (b);
      \draw[-, rounded corners=5pt] (ax2) -| (bd);
    \end{tikzpicture}}
    &
    \scalebox{0.75}{\begin{tikzpicture}
      % tensor
      \node at (-0.75, 0.75) (a) {$A$};
      \node at (0.75, 0.75) (b) {$B$};
      \node at (0, 0) (tens) {$\otimes$};
      \node at (0, -0.75) (ab) {$A \otimes B$};
      \draw[-] (a) -- (tens);
      \draw[-] (b) -- (tens);
      \draw[-] (tens) -- (ab);

      % par
      \node at (1.25, 0.75) (ad) {$A^\bot$};
      \node at (2.75, 0.75) (bd) {$B^\bot$};
      \node at (2, 0) (par) {$\parr_L$};
      \node at (2, -0.75) (adbd) {$A^\bot \parr B^\bot$};
      \draw[-] (ad) -- (par);
      \draw[-] (par) -- (adbd);
    \end{tikzpicture}}
    \\
    \hline
\end{tabular}
    \caption{The vehicle of a proof-structure and a test corresponding to a Danos-Regnier switching graph.}
    \label{fig:correction}
\end{figure}
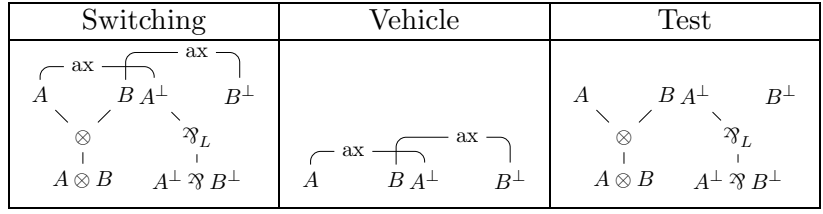

We call the translation of switching graphs \emph{tests}. They are defined in a very natural way by translating the nodes of the lower part of a proof-structure $\mathcal{S}$ into constellations:

\begin{itemize}
    %% Ax
    \item $A^\bigstar = [\tcol{-t.\mathtt{addr}_\mathcal{S}(A)}, \qray[+]{A}]$ where $A$ is a conclusion of axiom;
    %% Par
    \item $(A \parr_L B)^\bigstar = \gparl{A}{B}$;
    \item $(A \parr_R B)^\bigstar = \gparr{A}{B}$;
    %% Tensor
    \item $(A \otimes B)^\bigstar = \gtens{A}{B}$;
    \item We add $\gconc{A}$ for each conclusion $A$.
\end{itemize}

where $\tcol{-t.\mathtt{addr}_\mathcal{S}(A)}$ corresponds to the address of $A$ relatively to $\mathcal{S}$ (it is basically the same as the translation of an atom but with a different colour). The colour $\tcol{t}$ stands for "typing".

\begin{definition}[Logical correctness]
The translation of a proof-structure of conclusion $\vdash A_1, ..., A_n$ is said to be correct when for each translation of switching graph, their union normalises into $[p_{A_1}(x), ..., p_{A_n}(x)]$.
\end{definition}

The essential point of the translation is that the corresponding dependency graph, obtained by describing how the rays can be connected to each other, will have the same structure as a switching graph.

\begin{example}
Here is an example with the switching graph and the test of figure \ref{fig:correction}:

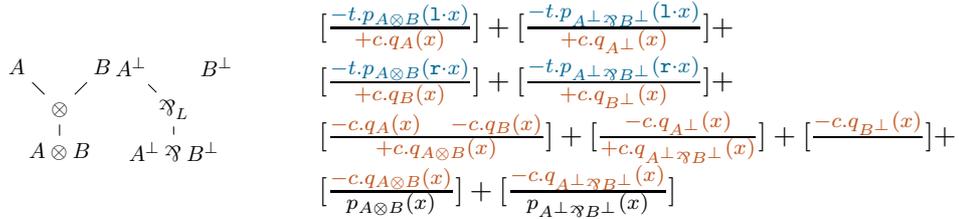
\begin{center}
\begin{minipage}{0.275\textwidth}
    \scalebox{0.75}{\begin{tikzpicture}
      % tensor
      \node at (-0.75, 0.75) (a) {$A$};
      \node at (0.75, 0.75) (b) {$B$};
      \node at (0, 0) (tens) {$\otimes$};
      \node at (0, -0.75) (ab) {$A \otimes B$};
      \draw[-] (a) -- (tens);
      \draw[-] (b) -- (tens);
      \draw[-] (tens) -- (ab);

      % par
      \node at (1.25, 0.75) (ad) {$A^\bot$};
      \node at (2.75, 0.75) (bd) {$B^\bot$};
      \node at (2, 0) (par) {$\parr_L$};
      \node at (2, -0.75) (adbd) {$A^\bot \parr B^\bot$};
      \draw[-] (ad) -- (par);
      \draw[-] (par) -- (adbd);
    \end{tikzpicture}}
\end{minipage}
\begin{minipage}{0.7\textwidth}
    $\dgloc{A \otimes B}{\mathtt{l} \cdot x}{A}+
     \dgloc{A^\bot \parr B^\bot}{\mathtt{l} \cdot x}{A^\bot}+$ \\
     $\dgloc{A \otimes B}{\mathtt{r} \cdot x}{B}+
     \dgloc{A^\bot \parr B^\bot}{\mathtt{r} \cdot x}{B^\bot}+$ \\
    $\dgtens{A}{B}{A \otimes B}+\dgparl{A^\bot}{B^\bot}{A^\bot \parr B^\bot}+$ \\
    $\dgconc{A \otimes B}+\dgconc{A^\bot \parr B^\bot}$
\end{minipage}
\end{center}

When connected to the vehicle: \[[\tlocus{A \otimes B}{\mathtt{l} \cdot x}, \tlocus{A^\bot \parr B^\bot}{\mathtt{l} \cdot x}]+[\tlocus{A \otimes B}{\mathtt{r} \cdot x}, \tlocus{A^\bot \parr B^\bot}{\mathtt{r} \cdot x}]\]

we obtain the following dependency graph:

\bigskip
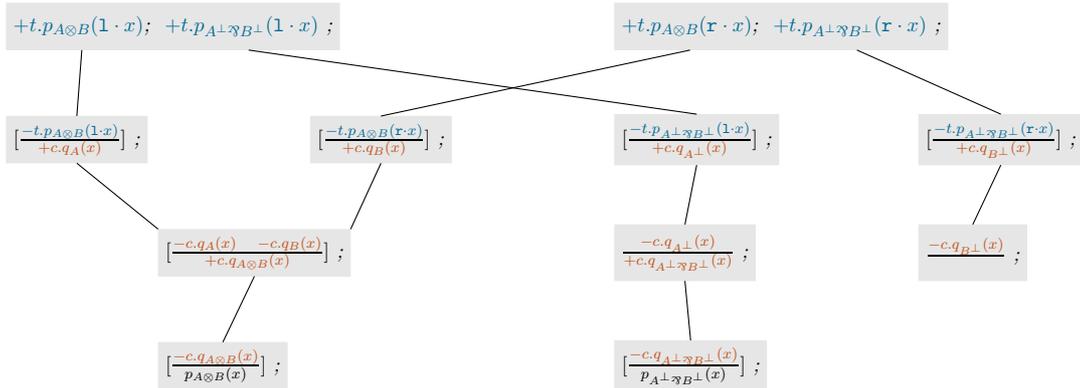
\begin{tikzpicture}[every node/.style={scale=0.7}]
    \tstar[25]{ax1}{(0, 0)}{{
        {$\tcol{+t.p_{A \otimes B}(\mathtt{l} \cdot x)}$},
        {$\tcol{+t.p_{A^\bot \parr B^\bot}(\mathtt{l} \cdot x)}$}
    }}
    \tstar[25]{ax2}{(8, 0)}{{
        {$\tcol{+t.p_{A \otimes B}(\mathtt{r} \cdot x)}$},
        {$\tcol{+t.p_{A^\bot \parr B^\bot}(\mathtt{r} \cdot x)}$}
    }}
    \tstar[25]{loc1}{(0, -1.5)}{{
        {$\dgloc{A \otimes B}{\mathtt{l} \cdot x}{A}$}
    }}
    \tstar[25]{loc2}{(4, -1.5)}{{
        {$\dgloc{A \otimes B}{\mathtt{r} \cdot x}{B}$}
    }}
    \tstar[25]{loc3}{(8, -1.5)}{{
        {$\dgloc{A^\bot \parr B^\bot}{\mathtt{l} \cdot x}{A^\bot}$}
    }}
    \tstar[25]{loc4}{(12, -1.5)}{{
        {$\dgloc{A^\bot \parr B^\bot}{\mathtt{r} \cdot x}{B^\bot}$}
    }}
    \tstar[25]{tens}{(2, -3)}{{
        {$\dgtens{A}{B}{A \otimes B}$}
    }}
    \tstar[30]{parl}{(8, -3)}{{
        {$\frac{\ccol{-c.q_{A^\bot}(x)}}{\ccol{+c.q_{A^\bot \parr B^\bot}(x)}}$}
    }}
    \tstar[30]{parr}{(12, -3)}{{
        {$\frac{\ccol{-c.q_{B^\bot}(x)}}{}$}
    }}
    \tstar[25]{c1}{(2, -4.5)}{{
        {$\dgconc{A \otimes B}$}
    }}
    \tstar[25]{c2}{(8, -4.5)}{{
        {$\dgconc{A^\bot \parr B^\bot}$}
    }}
    \draw (ax1/0.south) -- (loc1/0.north);
    \draw (ax1/1.south) -- (loc3/0.north);
    \draw (ax2/0.south) -- (loc2/0.north);
    \draw (ax2/1.south) -- (loc4/0.north);
    \draw (loc1/0.south) -- (tens/0.north west);
    \draw (loc2/0.south) -- (tens/0.north east);
    \draw (loc3/0.south) -- (parl/0.north);
    \draw (loc4/0.south) -- (parr/0.north);
    \draw (tens/0.south) -- (c1/0.north);
    \draw (parl/0.south) -- (c2/0.north);
\end{tikzpicture}

\bigskip
structurally corresponding to the following switching graph:

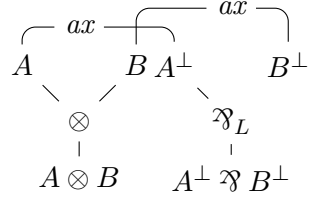
\begin{center}
\begin{tikzpicture}
  % tensor
  \node at (-0.75, 0.75) (a) {$A$};
  \node at (0.75, 0.75) (b) {$B$};
  \node at (0, 0) (tens) {$\otimes$};
  \node at (0, -0.75) (ab) {$A \otimes B$};
  \draw[-] (a) -- (tens);
  \draw[-] (b) -- (tens);
  \draw[-] (tens) -- (ab);

  % par
  \node at (1.25, 0.75) (ad) {$A^\bot$};
  \node at (2.75, 0.75) (bd) {$B^\bot$};
  \node at (2, 0) (par) {$\parr_L$};
  \node at (2, -0.75) (adbd) {$A^\bot \parr B^\bot$};
  \draw[-] (ad) -- (par);
  \draw[-] (par) -- (adbd);

  % ax
  \node at (0, 1.25) (ax1) {ax};
  \node at (2, 1.5) (ax2) {ax};
  \draw[-, rounded corners=5pt] (ax1) -| (a);
  \draw[-, rounded corners=5pt] (ax1) -| (ad);
  \draw[-, rounded corners=5pt] (ax2) -| (b);
  \draw[-, rounded corners=5pt] (ax2) -| (bd);
\end{tikzpicture}
\end{center}

Since the matching of the constellation is exact and that the dependency graph is a tree, only the free rays will be kept in the normal form. We obtain $[p_{A \otimes B}(x), p_{A^\bot \parr B^\bot}(x)]$.
\end{example}

\begin{example}
If we have the following test instead:

\begin{minipage}{0.2\textwidth}
    \scalebox{0.75}{\begin{tikzpicture}
      % tensor
      \node at (-0.75, 0.75) (a) {$A$};
      \node at (0.75, 0.75) (b) {$A^\bot$};
      \node at (0, 0) (tens) {$\otimes$};
      \node at (0, -0.75) (ab) {$A \otimes A^\bot$};
      \draw[-] (a) -- (tens);
      \draw[-] (b) -- (tens);
      \draw[-] (tens) -- (ab);
  \end{tikzpicture}}
\end{minipage}
\begin{minipage}{0.7\textwidth}
    $\dgloc{A \otimes A^\bot}{\mathtt{l} \cdot x}{A}+
     \dgloc{A \otimes A^\bot}{\mathtt{r} \cdot x}{A^\bot}+$ \\
    $\dgtens{A}{A^\bot}{A \otimes A^\bot}+\dgconc{A \otimes A^\bot}$
\end{minipage}

When connected to the vehicle $[\clocus{A \otimes A^\bot}{\mathtt{l} \cdot x}, \clocus{A \otimes A^\bot}{\mathtt{r} \cdot x}]$, a loop appears in the dependency graph and since the matching is exact, we can construct infinitely many correct diagrams. The constellation is not strongly normalising.
\end{example}

\subsection{What is a proof?}

We started from very general untyped objects which are the constellations and reconstructed the elementary bricks of multiplicative linear logic. Our framework liberalise proof-structures by naturally allowing all sorts of constructions:
\begin{itemize}
    \item an atomic pre-proof of conclusion $\vdash A$ as the constellation $\{[p_A(x)]\}$,
    \item an n-ary axiom as the constellation $\{[p_{A_1}(x), ..., p_{A_n}(x)]\}$.
\end{itemize}

\bigskip
We can finally define what the full translation of a proof-structure is. A proof-structure is translated into a triple $(\Phi_{V}, \Phi_{C}, {\color{MidnightBlue}\Phi_{F}})$ where:
\begin{itemize}
    \item $\Phi_{V}$ is the uncoloured vehicle made of translations of axioms,
    \item $\Phi_{C}$ is the uncoloured translation of cuts,
    \item ${\color{MidnightBlue}\Phi_{F}}$ is coloured translation of all ordeals (the format).
\end{itemize}

We see that proof-structures can be considered as being made of three components. This shows that proof-structures, although being considered untyped, actually come with an implicit typing corresponding to the bottom of the proof-structure: we made an implicit logical assumptions computationally explicit.

It also exhibits a confusion in proof-net theory where we manipulate hybrid objects containing a computational and logical part without the possibility of separating them. The GoI separates these two parts so that the computational and logical parts of a proof can be distinguished and studied independently. It is like studying $\lambda$-terms and types as separate entities rather than the single hybrid notion of typed $\lambda$-term. In particular, tests can live independently of any notion of proof since they are only generated by formulas. Moreover, any constellation can be tested with the tests coming from formulas.

\section{Two notions of type/formula}

Depending of whether types or programs come first, we have two distinct notions of typing which are reunited in the transcendental syntax. Girard talks about \emph{existentialism} when programs come first and \emph{essentialism} when types do.

The main required ingredient is an \emph{orthogonality relation} $\perp$ opposing constellations depending of what we consider a \emph{correct} interaction. This corresponds to the subjective side of logic: defining logic from a selected point of view on computational interaction. Several choices are possible and lead to different results. For instance, we may consider:
\begin{itemize}
    \item $\Phi \perp \Phi'$ when $|\exec(\Phi \uplus \Phi')| < \infty$ (strong normalisation) which captures MLL+MIX correctness.
    \item $\Phi \perp \Phi'$ when $|\exec(\Phi \uplus \Phi')| = 1$ which seems to capture MLL correctness but which is actually insufficient.
\end{itemize}

\subsection{Existentialist typing / Girard's use}

The first notion of typing, which I call \emph{interactive typing}, appears in ludics \cite{girard2001locus}, the GoI and in realisability interpretations \cite{krivine2009interactive, kleene1945interpretation, beffara2006concurrent}. Instead of types as syntactic labels, we consider types as collections of programs. Types are then seen as descriptions of computational behaviours or properties. With this idea, types are \emph{designed} instead of \emph{pre-defined}. It is called \emph{existentialist} because types (the essence) comes after the object.

By grouping some constellations together, it is possible to design various sort of types. In this section, we show how to design types corresponding to MLL formulas but atypical connectives can also be constructed (as the \emph{insinuation} connective of Girard \cite{girard2018logique}).

This notion of typing is based on the idea of \emph{meaning-as-use} where computational interaction defines the meaning of a group of constellations. Girard illustrates this with the (outdated) idea of DVD player. If you do not know what is a DVD player then you can either read a definition in the dictionary, which corresponds to a semantic explanations which are standard in logic. Another way is to define it by what it can interact with: DVD. When you put what you think are DVD, some interactions will go well and others will not (this is the orthogonality relation). But then, a DVD player can be defined by all the DVD it can read. In the same way, the notion of DVD can be defined by all DVD players which can read it.

\begin{description}
    \item[Pre-behaviour] A pre-behaviour is a set of constellations.
    \item[Orthogonal] Given a pre-behaviour $\mathbf{A}$, its orthogonal, written $\mathbf{A}^\bot$, is the set of all constellations which are strongly normalising when interacting with the constellations of $\mathbf{A}$. It corresponds to linear negation.
    \item[Behaviour] A \emph{behaviour} (or formula) $\mathbf{A}$ is the orthogonal of another pre-behaviour $\mathbf{B}$ i.e $\mathbf{A} = \mathbf{B}^\bot$. It means that it interacts well (with respects to the orthogonality) with another pre-behaviour. It is equivalent to say that $\mathbf{A} = \mathbf{A}^{\bot\bot}$ meaning that it is closed by interaction. Such pre-behaviours correspond to the ones which can be entirely characterised by tests (objects certifying membership in $\mathbf{A}$), in other words, be \emph{testable}.
    \item[Atoms] We define atoms with a \emph{basis of interpretation} $\Phi$ associating for each type variable $X_i$ a distinct behaviour $\Phi(X_i)$. It represents a choice of formula for each variable. A more satisfactory way to handle variables is to consider second-order quantification, in which case we need further correctness tests. Since our atoms are represented by rays (thus concrete entities), Girard even considers a constant $\fu$ (fu) \cite{girard2018logique} which is self-dual.
    \item[Tensor] The tensor $\mathbf{A} \otimes \mathbf{B} := \{\Phi_A \uplus \Phi_B \mid \Phi_A \in \mathbf{A}, \Phi_B \in \mathbf{B}\}^{\bot\bot}$ of two behaviours is constructed by pairing all the constellations of $\mathbf{A}$ with the ones of $\mathbf{B}$ by using a multiset union of constellations $\Phi_1 \uplus \Phi_2$. The behaviour $\mathbf{A}$ and $\mathbf{B}$ have to be disjoint in the sense that they cannot be connected together by two matching rays. Note that the cut is the same thing except taht the constellations can interact. We use the double orthogonal $(\cdot)^{\bot\bot}$ to ensure that we have a behaviour because it is not always the case depending on the relation $\perp$.
    \item[Par and linear implication] As usual in linear logic, the par and linear implication are defined from the tensor and the orthogonal: $A \parr B := (A^\bot \otimes B^\bot)^\bot$ and $A \multimap B := A^\bot \parr B$. Notice that these connectives are not simply arbitrary definitions on labels but that they have a computational interpretation. The constellations of $A \parr B$ are the one passing the tests of $(A^\bot \otimes B^\bot)$.
    \item[Alternative definition for linear implication] An alternative but equivalent definition of linear implication is the set of all constellations $\Phi$ such that if we put them against any constellation of $\mathbf{A}$, the execution produces a constellation of $\mathbf{B}$. This is a standard definition in realisability theory.
\end{description}

It is possible to design any type we want providing it is a behaviour. For instance, we could design a type $\mathbf{Cut}(\mathbf{A}, \mathbf{B})$ which is exactly $\mathbf{A} \otimes \mathbf{B}$ without requiring rays to be unconnectable, and have $\mathbf{A} \otimes \mathbf{B} \subseteq \mathbf{Cut}(\mathbf{A}, \mathbf{B})$.

\begin{example}[Acceptation in finite automata]
From the previous encoding of automata, we can observe a duality between automata and words. We say that $A^\bigstar \bot w^\bigstar$ whenever $\exec(A^\bigstar + w^\bigstar) \neq \emptyset$. An automaton becomes orthogonal to all the words it accepts and a word is orthogonal to all the automata which recognise it. Notice that if we have $L = \{w \mid w \text{ ends with } 00\}$, then $L^\bot$ without restriction is not exactly the set of automata recognising $L$: it contains more constellations which are also able to recognise $L$ in a different way or irrelevant constellations. Other more accurate orthogonality relations can filter constellations.

If we have an operation of pairwise concatenation $L_1 \bullet L_2 = \{w_1w_2 \mid w_i \in L_i\}^{\bot\bot}$ (the double orthogonal is here to ensure that we have a behaviour), then $(L_1 \bullet L_2)^\bot$ is indeed the set of constellations recognising $L_1 \bullet L_2$ but this also defines (by duality), an operator on automata $\mathbf{A_1} \circ \mathbf{A_1} = (\mathbf{A_1}^\bot \bullet \mathbf{A_2}^\bot)^\bot$ which construct automata recognising the concatenation of two languages. This is similar to Terui's works \cite{terui2011computational} which studied a computational variant of ludics.
\end{example}

\begin{example}[Queries and answers in logic programming]
Let \[\Phi_\nat^+ = [+add(0, y, y)]+[-add(x, y, z), +add(s(x), y, s(z))]\] be a constellation computing the sum of two natural numbers. We consider the strong normalisation of the union of two constellations as orthogonality. Let
\[\mathbf{Q_+} = \{[-add(s^n(0), s^m(0), r), r] \mid n, m \in \nat\}\] and $\mathbf{A_+} = \{[s^n(0)] \mid n \in \nat\}$. Take a constellation $[-add(s^n(0), s^m(0), r), r]$ and connect it with $\Phi_\nat^+$. All diagrams corresponding to $\Phi_\nat^+$ with $n$ occurrences of \[[+add(s(x), y, s(z)), -add(x, y, z)]\] can be reduced to a star $[s^{n+m}(0)]$. It is easy to check that all other diagram fails. Therefore, for all $\Phi \in \mathbf{Q_+}$, $\exec(\Phi \uplus \Phi_\nat^+) \in \mathbf{A_+}$ and $\Phi_\nat^+ \uplus \Phi$. If $\mathbf{Q_+}$ and $\mathbf{A_+}$ are behaviours (we need a more specific orthogonality) then $\Phi_\nat^+ \in \mathbf{Q_+} \multimap \mathbf{A_+}$. Although not explored here, it might be possible to retrieve existing type systems and extend them.
\end{example}

We can also imagine more interesting examples: typing for tilings models (thus typing for DNA computing), characterisation of properties of constellations (characterisation of complexity classes?).

Finally, by considering that meaning is defined by interaction, we got rid of semantics since there is no previously defined constraints about how formulas should be. Formulas can be defined as we wish by constructions between sets of constellations and a given point of view (orthogonality relation).

\subsection{Neo-essentialist typing / Girard's factory}

The problem with the previous interactive typing is that \emph{effective reasoning} is no more possible. This is the price for a semantic-free space. To take again the illustration of DVD player, you will never be able to tell if you have a DVD player or DVD in front of you unless you travel the world in the search of all the existing DVD players and DVD. But another way is possible: to have a sample of ``good'' candidates in order to proceed to a finite verification. It is what happens in factories when certifying a product (a car, a vaccine, a tool, a DVD, a DVD player). This is what Girard calls the \emph{usine} (\emph{factory} in English).

In terms of linear logic and constellations, imagine that we have a constellation $\Phi$ and that we would like to check if it is a proof of $\mathbf{A}$, hence checking if $\Phi \in \mathbf{A}$. Since it is a behaviour, we have to check if $\Phi$ passes the tests $\mathbf{B} := \mathbf{A}^\bot$ of $\mathbf{A}$, hence to check if $\Phi \in \mathbf{B}^\bot$. However, depending on $\perp$, the set of tests can be infinite which makes impossible the answer to ``do I have a proof of $\mathbf{A}$?''. The way to get out of this infinite hell is to materialise the correctness criterion of linear logic which provide a finite verification. For a formula label $A$, it is possible to define the pre-behaviour $\mathtt{Tests}(A)$ corresponding to the set of tests for $A$ such that $\mathtt{Tests}(A)^\bot \subseteq \mathbf{A}$. Hence, passing this finite battery of tests is sufficient to ensure membership to $\mathbf{A}$.

This notion of typing is close to the traditional typing of proof theory and type theory (Martin-Löf type theory for instance) which only needs finite verification (effective type checking by rules or algorithm). This traditional point of view handling types as syntactic labels is called \emph{essentialist typing} by Girard. However, these labels put constraints of the notion of types: some programs may not be typable (for instance $\lambda x.xx$ in simply typed $\lambda$-calculus). Types represent arbitrary constraints on computation.

Girard calls his factory typing a \emph{dessentialisation of types} since it retrieves finite verification again but in an open semantic-free space, without the constraints of type labels which are in his opinion, a pure reification of prejudices.

To sum-up the two notions of logical meaning, I quote Girard:
\begin{itemize}
    \item L'\emph{usine} enables to predict what proofs/programs will do.
    \item L'\emph{usage} is what proofs/programs actually do.
    \item Using mathematics, we have to show the accuracy of l'usine: how accurately it is able to certify the use/behaviour.
\end{itemize}

\subsection{A logical constant for atoms}

In transcendental syntax's fourth paper \cite{transyn4}, Girard gives hints for the definition of a self-dual behaviour corresponding to a new logical constant.

The idea is that in the stellar resolution, atoms are translated into concrete objects which are not substitutable, unlike in the original theory of proof-nets. It should be possible to group them in a behaviour corresponding to the type of atoms. But the only thing they share in common is their topology (a single point). In order to retrieve Girard's logical constant $\fu$, we will consider that two constellations sharing the same structure are seen as being in the same group in the point of view of typing.

This is actually an extension of the correctness criterion on partitions described before but adapted to the case of constellations \cite{transyn2}. We only give informal definitions.

A constellation $\Phi$ is said to be \emph{rooted} when it has at most one uncoloured ray for each of its stars. These uncoloured rays are called the \emph{roots} of $\Phi$, inducing a star $\mathtt{Roots}(\Phi)$ of the roots.

\begin{definition}[Orthogonality]
Two constellations $\Phi_1$ and $\Phi_2$ satisfying the two following conditions:
\begin{itemize}
    \item $|\pm\mathtt{Rays}(\Phi_1)| = |\pm\mathtt{Rays}(\Phi_2)|$ where $\pm\mathtt{Rays}$ returns the set of coloured rays of a constellation;
    \item exactly one of them is rooted;
\end{itemize}
are \emph{orthogonal}, written $\Phi_1 \perp \Phi_2$, if and only if $\exec(\Phi_1' \uplus \Phi_2') = \mathtt{Roots}(\Phi_i)$ where $\Phi_i$ with $i \in \{1,2\}$ is rooted, and $\Phi_1'$ and $\Phi_2'$ are $\Phi_1$ and $\Phi_2$ coloured with $\tcol{+t}$ and $\tcol{-t}$ (if there is already a colour for a ray, it is replaced, otherwise, if it is uncoloured, nothing changes).
\end{definition}

Now, remark that uncoloured rays do not participate in computation and simply corresponds to output. It is fair to consider that behaviours are equal up to removal of uncoloured rays but also up to change of terms providing the depency graphs has the same structure. We define an equivalence relation $\approx$ such that for two behaviours $\mathbf{A} \approx \mathbf{B}$ whenever for all $\Phi_A \in \mathbf{A}$ and $\Phi_B \in \mathbf{B}$, the dependency graph of $\Phi_A$ and $\Phi_B$ are isomorphic or $\mathbf{A}$ with all uncoloured rays removed is equal or equivalent (by $\approx$) to $\mathbf{B}$ with all uncoloured rays removed. We implicitly consider two behaviours equal when they are equivalent modulo $\approx$.

We can then define a behaviour corresponding to the type of atoms.

\begin{proposition}[Logical constant $\fu$]
The pre-behaviour $\fu = \{\{\phi\} \mid |\phi| = 1\}$ is a self-dual behaviour, that is $\fu = \fu^\bot = \fu^{\bot\bot}$.
\end{proposition}

Now, we can give ground to multiplicative propositions. All formula variables can be replaced by $\fu$. For instance $\vdash X_1^\bot \parr X_2^\bot, X_1 \otimes X_2$ can be translated into the behaviour $\vdash \fu \parr \fu, \fu \otimes \fu$.

There are few interesting points:
\begin{itemize}
    \item for any formula $A$, we can type atomic pre-proofs: $[\tcol{+t.p_A(x)}] \in \fu$;
    \item the axioms are now typable with $[\tcol{+t.p_A(x)}, \tcol{+t.p_{A^\bot}(x)}] \in \fu \parr \fu$;
    \item we can write stand-alone links: $[\tcol{+t.p_A(x)}, \tcol{+t.p_B(x)}] \in \fu \parr \fu$ and $[\tcol{+t.p_A(x)}]+[\tcol{+t.p_B(x)}] \in \fu \otimes \fu$.
\end{itemize}

One purpose of $\fu$ is to justify the axiom $\vdash A, A^\bot$ for a formula $A$ we know nothing about. We theoretically need an infinite verification since it has to work for any formula $A$. However, it is sufficient to check the cases $A:=\fu, A:=\fu \otimes \fu$ and $A:=\fu \parr \fu$ with $A^\bot:=\fu, A^\bot:=\fu \parr \fu, A^\bot:=\fu \otimes \fu$ (respectively), corresponding to the possible shapes of an axiom \cite{transyn4}. This is the three cases we would check if we would do an induction on $A$. For instance, assume that we have the two tests $A := \fu \parr \fu$ and $A^\bot := \fu \otimes \fu$. The orthogonal of these tests must be a constellation $[r_1, r_2]+[r_3, r_4]$ which is indeed isomorphic to the vehicle of a proof of $\vdash X_1^\bot \parr X_2^\bot, X_1 \otimes X_2$ with expansed axioms.

This orthogonality relation only cares about the \emph{shape} of constellations and their interactions. This new point of view of logic is what Girard calls \emph{morphologism} \cite{girard2018logique}.

\section{Extension with intuitionistic implication}

The exponentials have already been studied using flows (corresponding to binary stars) \cite{goi3, bagnol2014resolution} but the correctness was not considered yet until Girard's transcendental syntax \cite{transyn1}.

The idea is that the stellar resolution already has a built-in mechanism of duplication (by multiple matching and re-use of stars) and weakening (rejection of some unwanted diagrams).

We restrict the definitions to MLL extended with the intuitionistic implication $A \Rightarrow B := \oc A \multimap B$ and its dual $(A \Rightarrow B)^\bot$ in order to follow Girard's papers \cite{goi1}.

\begin{itemize}
    \item Non-linear formulas are written $\underline{A}$ (this corresponds to $\wn A$) and only appear at top-level (for conclusions and not subformulas).
    \item We define new connectives $A \ltimes B$ (corresponding to $\wn A \parr B$) and $A \olessthan B$ (corresponding to $\oc A \otimes B$) with the following links:
    \begin{center}
    \begin{tikzpicture}
      \node[dot] at (-0.75, 0.75) (a) {};
      \node[dot] at (0.75, 0.75) (b) {};
      \node at (0, 0) (tens) {$\ltimes$};
      \node[dot] at (0, -0.75) (ab) {};
      \draw[-stealth] (a) -- (tens);
      \draw[-stealth] (b) -- (tens);
      \draw[-stealth] (tens) -- (ab);
      \node at (0, -1.5) (label) {Exponential par};
    \end{tikzpicture}
    \hspace{3cm}
    \begin{tikzpicture}
      \node[dot] at (-0.75, 0.75) (a) {};
      \node[dot] at (0.75, 0.75) (b) {};
      \node at (0, 0) (tens) {$\olessthan$};
      \node[dot] at (0, -0.75) (ab) {};
      \draw[-stealth] (a) -- (tens);
      \draw[-stealth] (b) -- (tens);
      \draw[-stealth] (tens) -- (ab);
      \node at (0, -1.5) (label) {Exponential tensor};
    \end{tikzpicture}
    \end{center}
    \item Weakened (erased) atoms $X_i$ of address $t$ coming from a conclusion $\underline{A}$ are translated as a star \[[\ccol{+c.p_A(t \cdot y)}, -w_A(t \cdot y), +w_A(t \cdot y)]\] so that any cut connected to it forms a non-saturated diagram which will be rejected. This is basically a... black hole!
    \item Derelicted (linearised) atoms $X_i$ of address $t$ coming from a conclusion $\underline{A}$ are translated as a ray $p_A(t \cdot \mathtt{d})$.
    \item Contracted (duplicated) atoms $X_i, X_j$ (with $i \neq j$) of address $t$ coming from a conclusion $\underline{A}$ are translated into rays $p_A(t \cdot (\mathtt{l} \cdot y))$ and $p_A(t \cdot (\mathtt{r} \cdot y))$.
    \item Promoted atoms $X_i$ of address $t$ coming from a conclusion $A$ are translated into $p_A(t \cdot y)$ for a fresh variable $y$ and its auxiliary formulas $B_i$ into $p_{B_i}(t_i \cdot y)$ (representing the other parts of the associated exponential box).
\end{itemize}

We can remark that subterms of the form $t \cdot u$ are used. The term $t$ corresponds to the multiplicative part encoding the address of atoms as before and the term $u$ encodes the nested boxes (e.g $(t \cdot y_1) \cdot y_2$) and the copy identifier for the contraction (left and right copy). In the case of dereliction, we prevent any duplication on the current layer by a constant $\mathtt{d}$. This gives a box-free approach to exponentials where box dependency is simulated by matchability between terms.

To illustrate these notions, I present a simple $\lambda$-term. Because it looks like the least intuitive thing ever, it often makes me laugh but here is the identity function:
\[(\lambda x.x)^\bigstar := [\ccol{+c.p_{B^\bot \ltimes B}((\mathtt{l} \cdot x) \cdot \mathtt{d})}, \ccol{+c.p_{B^\bot \ltimes B}(\mathtt{r} \cdot x)}]\]
It is basically an axiom (hence a binary star) with a dereliction applied on the left atom so that the function applied (by cuts) to a box corresponding to an argument with addresses $t \cdot y$.

\subsection{The computational content of exponentials}

We illustrate the cut-elimination by few examples using the usual translation of simply typed $\lambda$-terms into proof-nets \cite{danosPHD, regnierPHD}.

\begin{example}[Case dereliction/box]
This corresponds to opening a box. We illustrate this case with the identity function applied to an argument: $(\lambda x.x)y$.

\bigskip
\begin{minipage}{0.35\textwidth}
\scalebox{0.65}{
\begin{tikzpicture}
    \node at (0, 4.5) (a1) {$B^\bot$};
    \node at (0, 3.5) (d) {$d$};
    \node at (0, 2.25) (a2) {$\underline{B}^\bot$};
    \node at (1.5, 2.25) (b2) {$B$};
    \node at (0.75, 1) (f2) {$\ltimes$};
    \node at (0.75, 0) (b) {$B^\bot \ltimes B$};
    \draw[-stealth] (f2) -- (b);
    \draw[-stealth] (a2) -- (f2);
    \draw[-stealth] (b2) -- (f2);
    \draw[-stealth] (a1) -- (d);
    \draw[-stealth] (d) -- (a2);

    \node at (3.75, 2.25) (c) {$B$};
    \node at (5.25, 2.25) (d) {$B^\bot$};
    \node at (4.5, 1) (tens) {$\olessthan$};
    \node at (4.5, 0) (cd) {$B \olessthan B^\bot$};
    \draw[-stealth] (c) -- (tens);
    \draw[-stealth] (d) -- (tens);
    \draw[-stealth] (tens) -- (cd);

    \node at (2.25, 2.25) (f1) {$B^\bot$};
    \node at (6.75, 2.25) (f2) {$B$};

    \node at (0.75, 5.5) (ax) {ax};
    \draw[-latex, rounded corners=5pt] (ax) -| (a1);
    \draw[-latex, rounded corners=5pt] (ax) -| (b2);

    \node at (2.5, -0.75) (cut) {cut};
    \draw[-latex, rounded corners=5pt] (b) |- (cut);
    \draw[-latex, rounded corners=5pt] (cd) |- (cut);
    \node at (3, 3) (ax) {ax};
    \draw[-latex, rounded corners=5pt] (ax) -| (f1);
    \draw[-latex, rounded corners=5pt] (ax) -| (c);
    \node at (6, 3) (ax) {ax};
    \draw[-latex, rounded corners=5pt] (ax) -| (f2);
    \draw[-latex, rounded corners=5pt] (ax) -| (d);
\end{tikzpicture}}
\end{minipage}
\begin{minipage}{0.6\textwidth}
It is translated into:
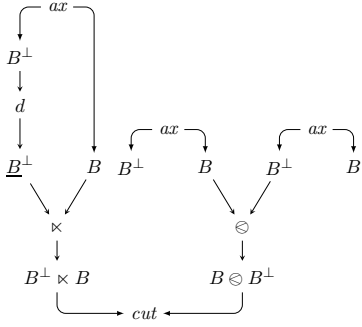
\[
[\ccol{+c.p_{B^\bot \ltimes B}((\mathtt{l} \cdot x) \cdot \mathtt{d})}, \ccol{+c.p_{B^\bot \ltimes B}(\mathtt{r} \cdot x)}]+
\]
\[
[\ccol{+c.p_{B^\bot}(x)}, \ccol{+c.p_{B \olessthan B^\bot}((\mathtt{l} \cdot x) \cdot y)}]+
\]
\[
[\ccol{+c.p_{B \olessthan B^\bot}(\mathtt{r} \cdot x)}, \ccol{+c.p_B(x)}]+
\]
\[
[\ccol{-c.p_{B^\bot \ltimes B}(x)}, \ccol{-c.p_{B \olessthan B^\bot}(x)}]
\]

It is similar to the multiplicative case. The essential point is that $\ccol{+c.p_{B^\bot \ltimes B}((\mathtt{l} \cdot x) \cdot \mathtt{d})}$ will interact with $\ccol{+c.p_{B \olessthan B^\bot}((\mathtt{l} \cdot x) \cdot y)}$ replacing $y$ by $\mathtt{d}$. It means that the rays corresponding to this box are no longer duplicable. In some sense, we opened the box.
\end{minipage}
\end{example}

\begin{example}[Case weakening/box]
It corresponds to a box erasure. We consider the simple example of right projection $(\lambda xy.y)z$.

\bigskip
\begin{minipage}{0.425\textwidth}
\scalebox{0.6}{
\begin{tikzpicture}
    \node at (-0.75, 1.5) (w) {$w$};
    \node at (0, 4.5) (a1) {$B^\bot$};
    \node at (0, 3.5) (d) {$d$};
    \node at (0, 2.25) (a2) {$\underline{B}^\bot$};
    \node at (1.5, 2.25) (b2) {$B$};
    \node at (0.75, 1) (f2) {$\ltimes$};
    \node at (-0.75, 0) (a) {$\underline{A}^\bot$};
    \node at (0.75, 0) (b) {$\underline{B}^\bot \ltimes B$};
    \node at (0, -1) (par) {$\ltimes$};
    \node at (0, -2) (ab) {$A^\bot \ltimes B^\bot \ltimes B$};
    \draw[-stealth] (a) -- (par);
    \draw[-stealth] (b) -- (par);
    \draw[-stealth] (par) -- (ab);
    \draw[-stealth] (w) -- (a);
    \draw[-stealth] (f2) -- (b);
    \draw[-stealth] (a2) -- (f2);
    \draw[-stealth] (b2) -- (f2);
    \draw[-stealth] (a1) -- (d);
    \draw[-stealth] (d) -- (a2);

    \node at (2.25, 0) (f1) {$A^\bot$};
    \node at (7.5, 0) (f2) {$(B \olessthan B^\bot)^\bot$};

    \node at (3.75, 0) (c) {$A$};
    \node at (5.25, 0) (d) {$B \olessthan B^\bot$};
    \node at (4.5, -1) (tens) {$\olessthan$};
    \node at (4.5, -2) (cd) {$A \olessthan B \olessthan B^\bot$};
    \draw[-stealth] (c) -- (tens);
    \draw[-stealth] (d) -- (tens);
    \draw[-stealth] (tens) -- (cd);

    \node at (2.25, -2.75) (cut) {cut};
    \draw[-latex, rounded corners=5pt] (ab) |- (cut);
    \draw[-latex, rounded corners=5pt] (cd) |- (cut);
    \node at (0.75, 5.5) (ax) {ax};
    \draw[-latex, rounded corners=5pt] (ax) -| (a1);
    \draw[-latex, rounded corners=5pt] (ax) -| (b2);
    \node at (3, 1.5) (ax) {ax};
    \draw[-latex, rounded corners=5pt] (ax) -| (f1);
    \draw[-latex, rounded corners=5pt] (ax) -| (c);
    \node at (6, 1.5) (ax) {ax};
    \draw[-latex, rounded corners=5pt] (ax) -| (f2);
    \draw[-latex, rounded corners=5pt] (ax) -| (d);
\end{tikzpicture}}
\end{minipage}
\begin{minipage}{0.5\textwidth}
It is translated into the constellation:
\[[\ccol{+c.p_{A^\bot \ltimes B^\bot \ltimes B}((\mathtt{l} \cdot x) \cdot y)},\]\[+w_{A^\bot \ltimes B^\bot \ltimes B}((\mathtt{l} \cdot x) \cdot y),\]
\[-w_{A^\bot \ltimes B^\bot \ltimes B}((\mathtt{l} \cdot x) \cdot y)]+\]
\[
[\ccol{+c.p_{A^\bot \ltimes B^\bot \ltimes B}((\mathtt{r} \cdot \mathtt{l} \cdot x) \cdot \mathtt{d})},\]
\[\ccol{+c.p_{A^\bot \ltimes B^\bot \ltimes B}(\mathtt{r} \cdot \mathtt{r} \cdot x)}]+
\]
\[
[\ccol{+c.p_{A^\bot}(x \cdot y)}, \ccol{+c.p_{A \olessthan B \olessthan B^\bot}((\mathtt{l} \cdot x) \cdot y)}]+
\]
\[
[\ccol{+c.p_{A \olessthan B \olessthan B^\bot}(\mathtt{r} \cdot x)}, \ccol{+c.p_{A \olessthan B \olessthan B^\bot}(x)}]+
\]
\[
[\ccol{-c.p_{A^\bot \ltimes B^\bot \ltimes B}(x)}, \ccol{-c.p_{A \olessthan B \olessthan B^\bot}(x)}]
\]
\end{minipage}

\bigskip
We can see that the cut will be duplicated into:
\[[\ccol{-c.p_{A^\bot \ltimes B^\bot \ltimes B}((\mathtt{l} \cdot x) \cdot y)},\ccol{-c.p_{A \olessthan B \olessthan B^\bot}((\mathtt{l} \cdot x) \cdot y)}]\]
and \[[\ccol{-c.p_{A^\bot \ltimes B^\bot \ltimes B}(\mathtt{r} \cdot x)}, \ccol{-c.p_{A \olessthan B \olessthan B^\bot}(\mathtt{r} \cdot x)}].\]

The ray $\ccol{-c.p_{A^\bot \ltimes B^\bot \ltimes B}((\mathtt{l} \cdot x) \cdot y)}$ will match the weakening star and all diagrams using terms matchable with the weakening star will be excluded (because no saturation is possible). This indeed corresponds to box erasure. Actually, it is also possible to put ``plugs'' $\ccol{+c.p_{A^\bot \ltimes B^\bot \ltimes B}(t \cdot y)}$ instead of the weakening star so that we obtain the empty star for each erased boxes. This is closer to cut-elimination of proof-nets which leaves some weakening nodes in the normal form.
\end{example}

\begin{example}[Case contraction/box]
It corresponds to a box duplication. Instead of illustrating it by the translation of a $\lambda$-term, we will describe how it works:
\[
[\ccol{+c.p_A(t \cdot (\mathtt{l} \cdot y))}, ...]+[\ccol{+c.p_A(t \cdot (\mathtt{r} \cdot y))}, ...]+[\ccol{+c.p_B(t \cdot y)}, ...]+[\ccol{-c.p_A(x)}, \ccol{-c.p_B(x)}]
\]

The cut star forces the matching between rays for $A$ and for $B$. We see that $\ccol{+c.p_B(t \cdot y)}$ can interact with both $\ccol{+c.p_A(t \cdot (\mathtt{l} \cdot y))}$ and $\ccol{+c.p_A(t \cdot (\mathtt{r} \cdot y))}$, hence the star containing $\ccol{+c.p_B(t \cdot y)}$ and all the stars connected to it will be naturally duplicated.
\end{example}

\subsection{The logical content of exponentials}

As usual with proof-nets, if we allow the MIX rule then the correctness criterion is straightforward. In this case, we only need to check acyclicity. The only difference with the multiplicative case is that we must check that the variables we use are handled correctly:
\begin{itemize}
    \item in subterms $x \cdot y$, the variables $x$ and $y$ must be different;
    \item the left part of a $\ltimes$ must be of shape $x \cdot y$.
\end{itemize}

If we reject the MIX rule, only the left part of $\ltimes$ which can be cancelled is problematic. We present new tests which extend the previous multiplicative tests:

\begin{itemize}
    %% Ax
    \item $A^\bigstar = \gaxex{\mathcal{S}}{A}$ where $A$ comes from the left part of $\ltimes$ or $\olessthan$;
    %% Tens 1
    \item $(A \olessthan_x B)^\bigstar = [\ccol{-c.q_A(x \cdot x)}, \ccol{-c.q_B(x)}, \ccol{+c.q_{A \olessthan B}(x)}]$;
    %% Tens 2
    \item $(A \olessthan_\mathtt{l} B)^\bigstar = [\ccol{-c.q_A(x \cdot \mathtt{l})}, \ccol{-c.q_B(x)}, \ccol{+c.q_{A \olessthan B}(x)}]$;
    %% Par
    \item $(A \ltimes_L B)^\bigstar = [\ccol{-c.q_A(x \cdot y)}, \ccol{+c.q_{A \ltimes B}(x \cdot y)}] +$ \\
    $[\ccol{-c.q_B(x \cdot y)}, +q_B(x), -q_B(x)] \quad\text{(cancelling)}$;
    \item $(A \ltimes_R B)^\bigstar = [\ccol{-c.q_B(x)}, \ccol{+c.q_{A \ltimes B}(x)}] + [\ccol{-c.q_A(x \cdot y)}]+[\ccol{-c.q_A(x' \cdot y')}] \\ (\text{unless } x = x')$;
    %% Conclusion
    \item $(\underline{A})^\bigstar = [\ccol{-c.q_A(x)}]$ where $\underline{A}$ is a conclusion.
\end{itemize}

The two tests for $\olessthan$ force the presence of different variables. The test $\ltimes_R$ only cancels the non-linear formulas. The star $[\ccol{-c.q_A(x' \cdot y')}]$ is only used when $x'$ can be instantiated with something different from the variable $x$. This is a technical hack which forces the $t$ in the $t \cdot u$ cancelled to be $x$. This actually uses a technical extension of stellar resolution called \emph{coherent constellations} introduced by Girard \cite{transyn2}.

Let us focus on the test $\ltimes_L$ which relies on the mechanisms of stellar resolution. Structural rules only appear on the left of $\ltimes$ and we need both acyclicity and connectedness. The idea is that any star connected to the star $[\ccol{-c.q_B(x \cdot y)}, +q_B(x), -q_B(x)]$ will be erased because of the loop making it impossible to produce a saturated diagram. Remark that we ask for this test to be cancelling (the interaction normalises into the empty constellation $\emptyset$). The test connects the conclusion of $\ltimes$ to its left part. The possible cases for $A \ltimes B$ are the following ones:

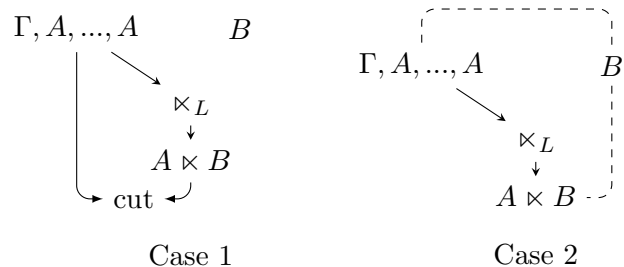
\begin{center}
\begin{minipage}{0.3\textwidth}
\begin{tikzpicture}
    \node at (-1.5, 1) (a) {$\Gamma, A, ..., A$};
    \node at (0.65, 1) (b) {$B$};
    \node at (0, 0) (tens) {$\ltimes_L$};
    \node at (0, -0.75) (ab) {$A \ltimes B$};
    \draw[-stealth] (a) -- (tens);
    \draw[-stealth] (tens) -- (ab);
    \node at (0, -2) (label) {Case 1};
    \node at (-0.75, -1.25) (cut) {cut};
    \draw[-latex, rounded corners=5pt] (a) |- (cut);
    \draw[-latex, rounded corners=5pt] (ab) |- (cut);
\end{tikzpicture}
\end{minipage}
\begin{minipage}{0.3\textwidth}
\begin{tikzpicture}
    \node at (-1.5, 1) (a) {$\Gamma, A, ..., A$};
    \node at (1, 1) (b) {$B$};
    \node at (0, 0) (tens) {$\ltimes_L$};
    \node at (0, -0.75) (ab) {$A \ltimes B$};
    \draw[-stealth] (a) -- (tens);
    \draw[-stealth] (tens) -- (ab);
    \node at (0, -1.5) (label) {Case 2};
    \draw[-, dashed, rounded corners=5pt] (a.north) |- (0, 1.75) -| (b);
    \draw[-, dashed, rounded corners=5pt] (b) |- (ab);
\end{tikzpicture}
\end{minipage}
\end{center}

\begin{description}
    \item[Case 1] The star corresponding to $\ltimes_L$ is connected to all (possible none) the copies of the atoms coming from $A$. Assume one of these atoms has a path leading to $A \ltimes B$ by a cut. Hence, we obtain a cycle and infinitely many correct diagrams. The constellation is not strongly normalising.
    \item[Case 2] Assume we do not have the cycle of case 1 and that some atoms of $A$ possibly have a path reaching an atom of $B$. If we wish to keep both connectivity, then all stars must be connected. Because of the star $[\ccol{-c.q_B(x \cdot y)}, +q_B(x), -q_B(x)]$, all stars connected to an atom of $B$ (actually the whole constellation) will be erased because unable to produce a saturated diagram. The normal form is $\emptyset$.
\end{description}

We then ask that for each tests (except the ones containing $\ltimes_L$ which should be cancelled) we get the star of conclusions (as usual) but we also require that we only get the non-linear conclusions in particular. Hence, for a proof of $\vdash \Gamma, \underline{\Delta}$, we should obtain the star of conclusions coming from $\Gamma$. This is due to two reasons:
\begin{itemize}
    \item there is a recurrent problem (not necessarily a flaw) of GoI: it is impossible to act on the auxiliary conclusions of boxes and to always preserve the paths of proofs with non-linear conclusions. Since we do not consider full exponentials, non-linear conclusions $\underline{A}$ cannot appear as conclusion of a proof anyway;
    \item morally, when performing a test and looking at the interaction between a vehicle and a test, it is impossible to guess how many times each conclusion appears in the normal form (if we use the orthogonality relation used for the definition of $\fu$). We could only test for the presence of a single star in the normal form but this is not sufficient because we cannot ensure that we got exactly the right conclusions.
\end{itemize}

\begin{example}[Correctness of identity function]
We check the correctness of the $\lambda$-term $\lambda x.x$.

\bigskip
\scalebox{0.85}
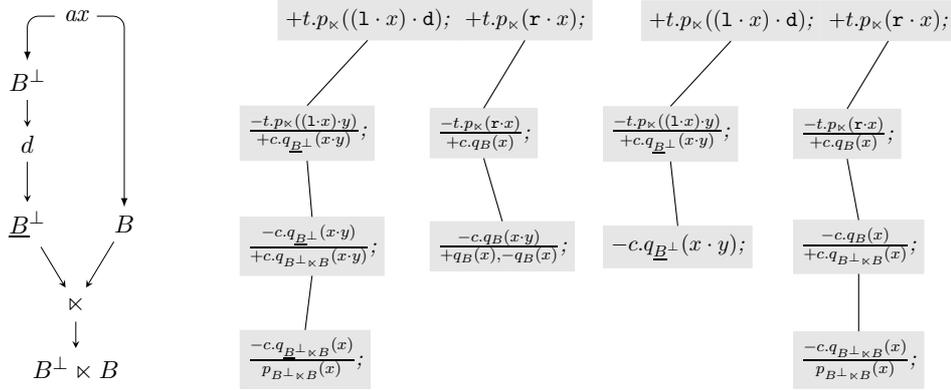
{
\begin{tikzpicture}
    \node at (0, 4.5) (a1) {$B^\bot$};
    \node at (0, 3.5) (d) {$d$};
    \node at (0, 2.25) (a2) {$\underline{B}^\bot$};
    \node at (1.5, 2.25) (b2) {$B$};
    \node at (0.75, 1) (f2) {$\ltimes$};
    \node at (0.75, 0) (b) {$B^\bot \ltimes B$};
    \draw[-stealth] (f2) -- (b);
    \draw[-stealth] (a2) -- (f2);
    \draw[-stealth] (b2) -- (f2);
    \draw[-stealth] (a1) -- (d);
    \draw[-stealth] (d) -- (a2);

    \node at (0.75, 5.5) (ax) {ax};
    \draw[-latex, rounded corners=5pt] (ax) -| (a1);
    \draw[-latex, rounded corners=5pt] (ax) -| (b2);
\end{tikzpicture}}
\hspace{1cm}
\begin{tikzpicture}[every node/.style={scale=0.75}]
    % vehicle
    \bistar{par}{(2.5, 0)}{$+t.p_{\ltimes}((\mathtt{l} \cdot x) \cdot \mathtt{d})$}{$+t.p_{\ltimes}(\mathtt{r}\cdot x)$}

    % gabarit
    \unistar{a2}{(2, -1.5)}{$\frac{-t.p_{\ltimes}((\mathtt{l} \cdot x) \cdot y)}{+c.q_{\underline{B}^\bot}(x \cdot y)}$}
    \unistar{a3}{(4.5, -1.5)}{$\frac{-t.p_{\ltimes}(\mathtt{r}\cdot x)}{+c.q_{B}(x)}$}

    \unistar{p1}{(2, -3)}{$\frac{-c.q_{\underline{B}^\bot}(x \cdot y)}{+c.q_{B^\bot \ltimes B}(x \cdot y)}$}

    \unistar{p1l}{(4.5, -3)}{$\frac{-c.q_{B}(x \cdot y)}{+q_B(x), -q_B(x)}$}

    \unistar{c1}{(2, -4.5)}{$\frac{-c.q_{\underline{B}^\bot \ltimes B}(x)}{p_{B^\bot \ltimes B}(x)}$}

    \draw (par/0.south) -- (a2/0.north);
    \draw (par/1.south) -- (a3/0.north);

    \draw (a3/0.south) -- (p1l/0.north);

    \draw (a2/0.south) -- (p1/0.north);
    \draw (p1/0.south) -- (c1/0.north);
\end{tikzpicture}
\begin{tikzpicture}[every node/.style={scale=0.75}]
    % vehicle
    \bistar{par}{(2.5, 0)}{$+t.p_{\ltimes}((\mathtt{l} \cdot x) \cdot \mathtt{d})$}{$+t.p_{\ltimes}(\mathtt{r} \cdot x)$}

    % gabarit
    \unistar{a2}{(2, -1.5)}{$\frac{-t.p_{\ltimes}((\mathtt{l} \cdot x) \cdot y)}{+c.q_{\underline{B}^\bot}(x \cdot y)}$}
    \unistar{a3}{(4.5, -1.5)}{$\frac{-t.p_{\ltimes}(\mathtt{r} \cdot x)}{+c.q_{B}(x)}$}

    \unistar{p1l}{(2, -3)}{$-c.q_{\underline{B}^\bot}(x \cdot y)$}
    \unistar{p1}{(4.5, -3)}{$\frac{-c.q_{B}(x)}{+c.q_{B^\bot \ltimes B}(x)}$}

    \unistar{c1}{(4.5, -4.5)}{$\frac{-c.q_{B^\bot \ltimes B}(x)}{p_{B^\bot \ltimes B}(x)}$}

    \draw (par/0.south) -- (a2/0.north);
    \draw (par/1.south) -- (a3/0.north);

    \draw (a3/0.south) -- (p1/0.north);
    \draw (a2/0.south) -- (p1l/0.north);
    \draw (p1/0.south) -- (c1/0.north);
\end{tikzpicture}

\medskip
We obtain two diagrams corresponding to the two possible correction graphs. The left one using $\ltimes_L$ is erased. The right one using $\ltimes_R$ normalises into $[p_{B^\bot \ltimes B}(x)]$. Hence, the identity function is logically correct.
\end{example}

\subsection{Expansionals connectives}

Using a similar idea, Girard suggested to construct new connectives called \emph{expansionals} which works like exponentials but on linear rays instead. It is a weak form of exponentials. We define two connectives $\uppar A$ and $\downtens A$, which only exist in combination with $\otimes$ and $\parr$ as in $\olessthan$ and $\ltimes$, defined by the following tests:

\begin{itemize}
    %% Tens 1
    \item $(\downtens_\mathtt{f} A \otimes B)^\bigstar = [\ccol{-c.q_A(f(x))}, \ccol{-c.q_B(x)}, \ccol{+c.q_{\downtens A \otimes B}(x)}]$;
    %% Tens 2
    \item $(\downtens_\mathtt{g} A \otimes B)^\bigstar = [\ccol{-c.q_A(g(x))}, \ccol{-c.q_B(x)}, \ccol{+c.q_{\downtens A \otimes B}(x)}]$;
    %% Par
    \item $(\uppar_L A \parr B)^\bigstar = [\ccol{-c.q_A(x)}, \ccol{+c.q_{\uppar A \parr B}(x)}] + [\ccol{-c.q_B(x)}, +q_B(x), -q_B(x)]$ \\ (cancelling);
    \item $(\uppar_R A \parr B)^\bigstar = [\ccol{-c.q_B(x)}, \ccol{+c.q_{\uppar A \ltimes B}(x)}] + [\ccol{-c.q_A(x)}]$.
\end{itemize}

It is then possible to create a behaviour for a new implication called \emph{insinuation} defined by $A \rightarrowtail B := \downtens A \multimap B$. The two first tests ensures that we have a linear ray $q_A(x)$. The switching $\uppar_R$ is exactly the same as for $\parr_R$ and $\uppar_L$ is very similar to $\ltimes_L$. By this similarity, it is possible to erase formulas on the left of $\uparrow A \parr B$ and $A \rightarrowtail B$ hence has the behaviour of an affine implication. But more than that, Girard remarked that it could still do a weak contraction on atoms \cite{girard2018logique}.

\section{New horizons for second-order logic}

A novelty of the transcendental syntax is the computational reconstruction of predicate calculus but also a redefinition of first and second-order logic.

\subsection{Girard's first and second-order}

Girard explained in a paper written in French \cite{girard2018logique} that second-order logic is actually implicit in a lot of definitions of logic. For instance, the rule $(\lor e)$ for NJ implicitly requires a generic formula $C$ which assumes the existence of a given specific space of allowed formulas:
\begin{center}
\begin{prooftree}
    \hypo{}
    \ellipsis{}{A \lor B}
    \hypo{[A]}
    \ellipsis{}{C}
    \hypo{[B]}
    \ellipsis{}{C}
    \infer3[$\lor e$]{C}
\end{prooftree}
\qquad\qquad
\begin{prooftree}
    \hypo{A}
    \hypo{B}
    \infer2[$\land i$]{A \land B}
\end{prooftree}
\end{center}

In theorems of propositional logic such as $A \Rightarrow B \Rightarrow A$, the formulas $A$ and $B$ are implicitly universally quantified and should be understood as $\forall A\  B.\  A \Rightarrow B \Rightarrow A$. The same thing occurs for $P$ which is external in $\forall x. P(x)$. This is how Girard justifies the treatment of the additives and exponentials in second-order for proof-nets because they use rules which are morally of second-order.

If we look at the rule $\land i$, it only locally reunites hypotheses without any genericity or need of a system. Since MLL proof-nets are hypergraphs linking formulas, it is similar. Even the atoms are concrete objects: simple vertices with non-essential labels. First-order logic hence corresponds to the part of logic which uses nothing more than ``what is already here" and second-order logic as the part of logic referring to the set of all propositions (externally structured). This structuration of the space of all formulas is what Girard calls \emph{epidictic architecture}. By this point of view on second-order logic, it happens that predicate calculus is actually part of second-order logic while MLL extended with $\Rightarrow$ and $\fu$ is purely first-order. We have the new following distinction between first and second-order:
\begin{description}
    \item[First-order] $\otimes, \parr, \multimap, \fu, \wo, \ltimes, \olessthan, \Rightarrow$;
    \item[Second-order] $\wn, \oc, \oplus, \&, \forall, \exists, \mathbf{1}, \botb, \mathbf{0}, \topb$.
\end{description}

where the additives, exponentials and neutral elements (not detailled here) are defined as follows:
\[\mathbf{A} \& \mathbf{B} := \exists X.\  \oc (X \multimap \mathbf{A}) \otimes \oc (X \multimap \mathbf{B}) \otimes X,\]
\[\mathbf{A} \oplus \mathbf{B} := \forall X.\  (\mathbf{A} \multimap X) \Rightarrow (\mathbf{B} \multimap X) \Rightarrow X,\]
\[\oc\mathbf{A} := \forall X. (\mathbf{A} \Rightarrow X) \multimap X, \qquad \wn\mathbf{A} := (\oc\mathbf{A})^\bot,\]
\[\mathbf{1} := \oc\topb, \qquad \botb := \wn\mathbf{0}, \qquad \mathbf{0} = \forall X.X, \qquad \topb = \exists X.X.\]
These new formulations for additives come from considerations of the rules for $\land$ and $\lor$ in $\mathbf{NJ}$ which are morally of second-order.

According to Girard, there are two status of the notion of (logical) \emph{system}:
\begin{itemize}
    \item it is an \emph{unavoidable evil}. All systems are bad but we still need one that we will improve \cite{transyn4}. This is the case of second-order logic where doubt exist but we have more expressivity and complex structurations.
    \item Systems are useless. This is what Girard calls \emph{anarchy}. Certainty is absolute and structuration purely emerge from computation. This is only possible when considering first-order connectives.
\end{itemize}

\subsection{Girard's derealism}

The treatment of interactive typing (the \emph{usage}) for second-order linear logic is straightforward. Universal and existential quantification respectively correspond to an infinite intersection and union of behaviours where $X$ is a bound variable appearing in behaviours and $\mathbf{T}$ is any behaviour:
\[\forall X.\mathbf{A} := \bigcap_{\mathbf{T} \in \mathfrak{E}} \mathbf{A}\{X := \mathbf{T}\}
\qquad\qquad
\exists X.\mathbf{A} := \bigg(\bigcup_{\mathbf{T} \in \mathfrak{E}} \mathbf{A}\{X:=\mathbf{T}\}\bigg)^{\bot\bot} \]
where $\mathfrak{E}$ is an epidictic architecture defined as a set of behaviours closed by some chosen connectives. Notice the plain intersection for $\forall$ because it is closed by bi-orthogonal while it is not the case for $\exists$ which has the same restriction as for $\otimes$.

Now, the problem of logical correctness is less straightforward (the \emph{usine}). Girard's \emph{derealism} corresponds to a new status for proofs, coming from a new treatment of second-order quantification. His intuition comes from the second-order definition of natural numbers:
\[\mathbf{nat} := \forall X. (X \multimap X) \Rightarrow (X \multimap X).\]
He claims \cite{girard2018logique} that $\mathbf{nat}^\bot$ corresponds to iteration/induction on natural numbers and that if testing for $\mathbf{nat}^\bot$ was finite, we could determine which iterations are licit, which is problematic. Further details are needed in order to make this idea explicit but it is not investigated here. Girard has the intuition that reasoning should be finite, hence we should preserve a finite testing. This leads to quantified entities being part of the proof itself. 
\[\infer{\exists X.A}{A\{X:=T\}}\]
More generally, the problem is that in a formula $\exists X.A$, it is impossible to finitely ``guess'' the right existential witness, which would be necessary to define tests from a formula. But, nonetheless, the witness $T$ is actually hidden in the proof itself, hence part of the vehicle (as you can see in the rule of second-order existential quantification). In Girard's derealism, the vehicle comes together with an auxliary test for the existential witness.

A proof is now a tuple $(\Phi_V \uplus \Phi_M, \Phi_C, \Phi_F)$ where $\Phi_M$ is called a \emph{mould} and corresponds to the tests associated to an existential witness.

Since $X$ in $\exists X.A$ may appear positively or negatively ($X$ or $X^\bot$), the mold comes in two versions and we are interested in a balance between them (are they truly orthogonal to each other?) which is not discussed here but the reader can find more details in the fourth article of transcendental syntax \cite{transyn4}.

As for the universal quantification $\forall X.A$, in the case of MLL proof-nets, it is sufficient to consider the three cases $X:=\fu, X:=\fu\parr\fu$ and $X:=\fu\otimes\fu$ on the top of tests for $A$.

For technical reasons and to add more combinatorial complexity to proofs, we distinguish two classes of rays:
\begin{itemize}
    \item \emph{objective} rays which are the usual rays;
    \item \emph{subjetive} rays allowing internal colours, for instance $+a(-b(x))$. As a consequence, we can to consider a subjective logical constant, similar to $\fu$. Girard calls it $\wo$ (wo). It can be tested with the test $[\tcol{-t.p_\wo(t \cdot x)}, p_\wo(x)]$ where $t$ is subjective (e.g $+a(x)$) ensuring the presence of an internal colour.
\end{itemize}

A star is \emph{objective} if it only has objective rays and \emph{subjective} if it only has subjective rays. Otherwise, it is \emph{animist} (a mix of objective and subjective). Animist stars typically appear in behaviours such as $(\fu \otimes \wo)^\bot = \fu \parr \wo$. A constellation of $\fu \otimes \wo$ is of the shape $[r]+[r']$ where $r'$ is subjective but in the dual, we obtain $[r, r']$, hence an animist star. An \emph{épure} is a constellation without animist star, i.e. it can be put into the form $\Phi_O \uplus \Phi_S$ where $\Phi_O$ only contains objective stars and $\Phi_S$ subjective ones. This makes the idea of correct proof clearer: it has to be made of an épure with an objective part only containing binary stars (they represent axioms).

The interesting feature of this new combinatorics is that the status of stars (objective/subjective/animist) can change during the execution. Another consequence is that we can design a behaviour $\mathbf{0}$ containing a constellation which can interact with other constellations but which is not correct (because of animist stars). This gives this constant a computational content, something which is usually not considered in usual logic ($\mathbf{0}$ is usually considered empty because it has no proof).

Finally, second-order logic in the transcendental syntax is dependent of a system (called \emph{epidictic architecture} by Girard) which specify the shape of existential quantifiers or the possible shapes the values a variable can take in the tests for universal quantification. For MLL, it is sufficient to restrict the epidictic to MLL formulas but in general we can go beyond that since we have an \emph{open system}.

\subsection{Additive neutrals}

We illustrate the mechanisms of witnesses and subjective rays with additive neutrals by following the third paper of transcendental syntax \cite{transyn3}. We define the following rays:
\[C_\top(x) := p_\top(\mathtt{c} \cdot x) \qquad L_\top(x) := p_\top(\tcol{-t(\mathtt{l} \cdot x)}) \qquad R_\top(x) := p_\top(\tcol{-t(\mathtt{r} \cdot x)})\]

The neutral element $\top$ is defined by the orthogonal of the following tests where the second one is cancelling (as for $\ltimes_L$ in the exponentials):
\[
\top_1 : \iostar{\tcol{-t.C_\top(x)}, \tcol{-t.L_\top(x)}}{} + \iostar{\tcol{-t.R_\top(x)}}{p_\top(x)}
\qquad
\top_2 : \iostar{\tcol{-t.C_\top(x)}, \tcol{-t.L_\top(x)}}{p_\top(x)} \quad \text{(cancel)}
\]

The constellation $[\tcol{+t.C_\top(x)}] + [\tcol{+t.L_\top(x)}, \tcol{+t.R_\top(x)}]$ is an épure passing the tests. However, it is not a correct proof because the objective part is unary and not binary.

If we take $[\tcol{+t.C_\top(\mathtt{l} \cdot x)}, \tcol{+t.C_\top(\mathtt{r} \cdot x)}] + [\tcol{+t.L_\top(\mathtt{l} \cdot x)}] + [\tcol{+t.L_\top(\mathtt{r} \cdot x)}, \tcol{+t.R_\top(x)}]$ instead, then it is a correct and also passes the test. In fact, it is possible to check that for any $\Phi \in \topb$, we have $\Phi \in\  \wn(\fu \otimes \wo) \parr \wo = (\fu \otimes \wo) \ltimes \wo = (\fu \parr \wo) \Rightarrow \wo$ because of the possibility of duplicating $C_\top$ and $L_\top$ which preserves the fact of passing the tests.

\bigskip
We define the following rays:
\[C_\mathbf{0}(x) := p_\mathbf{0}(\mathtt{c} \cdot x) \qquad L_\mathbf{0}(x) := p_\mathbf{0}(\tcol{-t(\mathtt{l} \cdot x)}) \qquad R_\mathbf{0}(x) := p_\mathbf{0}(\tcol{-t(\mathtt{r} \cdot x)})\]
The neutral element $\mathbf{0}$ is defined from the following tests:
\[\mathbf{0}_1 : \iostar{\tcol{-t.C_\mathbf{0}(x)}}{} + \iostar{\tcol{-t.L_\mathbf{0}(x)}, \tcol{-t.R_\mathbf{0}(x)}}{p_\mathbf{0}(x)}
\qquad
\mathbf{0}_2 : \iostar{\tcol{-t.L_\mathbf{0}(x)}}{} + \iostar{\tcol{-t.C_\mathbf{0}(x)}, \tcol{-t.R_\mathbf{0}(x)}}{p_\mathbf{0}(x)}\]
\[\mathbf{0}_3 : \iostar{\tcol{-t.R_\mathbf{0}(x)}}{p_\mathbf{0}(x)}\]

The constellation $[\tcol{+t.R_\mathbf{0}(x)}]+[\tcol{+t.C_\mathbf{0}(x)}, \tcol{+t.L_\mathbf{0}(x)}]$ (containing an animist star) passes the tests but there is no épure. The tests are actually designed in order to be orthogonal to a constellation with animist stars (corresponding to a logically incorrect constellation). This new combinatorics for proofs allows to speak about \emph{logical coherence} (that $\mathbf{0}$ has no correct inhabitant). Since $\mathbf{0} = \topb^\bot$, we should have that for any $\Phi \in \mathbf{0}$, we have $\Phi \in (\fu \parr \wo) \olessthan \wo$ (by following the expression of $\topb$ in terms of $\fu$ and $\wo$).

\subsection{Predicate calculus}

Because of Girard's new distinction between first and second-order, I choose to use the term "predicate calculus" instead of "first-order logic" to avoid confusion.

As remarked before, the predicate calculus is part of second-order logic but few conceptual choices remain. Girard remarks that by looking at Leibniz's equality defining $a=b$ as $\forall X.X(a) \Leftrightarrow X(b)$, the predicate $X$ would play no role if written as a proof-net. We can connect $a$ and $b$ directly and $X$ is seen as a sort of modality restricting connexions.

Girard's idea is to use second-order logic with quantification restricted to multiplicative formulas. Hence, terms/individuals and predicates are encoded with multiplicative formulas and equality becomes linear equivalence $A \equiv B$ defined as $(A \multimap B) \otimes (B \multimap A)$. Such terms can typically be multiplicative combinations of $\fu$ such as $\fu \otimes (\fu \parr \fu)$. The encodings can be freely chosen but for individuals, we need to ensure technical properties such as the injectivity of the encoding in order to have that $f(t)=f(u)$ implies $t=u$. These ideas are presented in Girard's third article on Transcendental Syntax \cite{transyn3}.

We can define the following pairing of terms where $\mathbf{T}$ and $\mathbf{U}$ are propositions corresponding to some terms:
\[<\mathbf{T}, \mathbf{U}>^\bigstar := (\mathbf{T} \parr \mathbf{U}) \otimes (\mathbf{T} \parr \mathbf{T} \parr \mathbf{U})\]
Such a pairing can represent a function applied to a term as in the previous term $f(t)$. If we have two pairs $<\mathbf{T}, \mathbf{U}>$ and $<\mathbf{T}', \mathbf{U}'>$ it is possible to verify their equality by checking if $<\mathbf{T}, \mathbf{U}> \equiv <\mathbf{T}', \mathbf{U}'>$ is provable (contains a correct constellation). This is equivalent to checking if
\[(\mathbf{T} \parr \mathbf{U}) \otimes (\mathbf{T} \parr \mathbf{T} \parr \mathbf{U}) \equiv (\mathbf{T}' \parr \mathbf{U}') \otimes (\mathbf{T}' \parr \mathbf{T}' \parr \mathbf{U}')\]
is provable. Now assume that it is provable. If we think in terms of sequent calculus, if it is provable then $\mathbf{T} \equiv \mathbf{T}'$ and $\mathbf{U} \equiv \mathbf{U}'$ must be provable as well, with satisfies our requirement of injectivity.

In terms of epidictic (the structure of quantified formulas and existential witnesses), since the quantification is restricted to individuals and that individuals are encoded as multiplicative formulas, it is sufficient to quantify only over multiplicative formulas.

\subsection{System-free Peano arithmetic}

In the fourth article of transcendental syntax \cite{transyn4}, Girard suggests an encoding of integers by multiplicative combinations of $\fu$ and $\wo$. We define $\fu_n$ as:
\begin{itemize}
    \item $\bigotimes_{k=1}^n \fu$ when $n > 0$ and
    \item $\bigparr_{k=1}^{2-n} \fu$ when $n < 2$.
\end{itemize}
And we define $\wo_n$ by induction on $n$:
\[\wo_0 := \wo \qquad \wo_n := \fu_n \otimes \wo \text{ when } n \neq 0\]

Integers are defined as follows:
\[\overline{n} := \wo_n \text{ when $n$ is positive and } \overline{n} := \wo_n^\bot \text{ otherwise}.\]

For instance, $\overline{0} := \wo, \overline{3} := \fu \otimes \fu \otimes \fu \otimes \wo$ and $\overline{-3} = \fu \parr \fu \parr \fu \parr \wo$. An example of proof of $\overline{3}$ is the constellation $[+n_3(x)]+[+n_2(x)]+[+n_1(x)]+[+n_0(+z(x))]$ and an example of proof of $\overline{-3}$ is the constellation $[+n_3(x), -n_2(x), -n_1(x), -n_0(+z(x)), x]$.

It is possible to interpret arithmetic operations by logical connectives for two integers $n$ and $m$:
\[\overline{n+m} := \overline{n} \otimes \overline{m} \qquad \overline{n-m} := \overline{n} \multimap \overline{m} \qquad \overline{m = n} := (\overline{n} \multimap \overline{m}) \otimes (\overline{m} \multimap \overline{n}).\]

A novelty suggested by Girard \cite{transyn4} is to extend these definitions by adding variables and quantifiers so that we can reconstruct Peano Arithmetic. It is actually system-free in the sense that we reconstruct arithmetic in a larger system where new connectives or operations can be defined as we wish without the need for axioms. Axioms are actually proven in some sense.

\section{Visibility and non-classical truth}

In technical terms, it is usually considered that truth is a property invariant by cut-elimination which has a distinguished value (of falsity) for contradiction ($\mathbf{0}$ in linear logic). Girard shows that, in the transcendental syntax, it is possible to define non-classical notions of truth which invalidate an idea of unique and absolute truth.

His notion of truth is called \emph{visibility}. What is invisible is what we do not want to see but which are essential in the functioning of the system (think of hidden files which are used by some programs but invisible for the user). In particular, an ambition of the transcendental syntax is to exhibit those ``hidden files of logic".

The idea is to take inspiration from Girard's Blind Spot \cite{blindspot} (end of first volume) where appears the \emph{Euler-Poincaré invariant} which returns $1$ for trees. It is a necessary condition for trees but not a sufficient one, hence not sufficient to reformulate the Danos-Regnier correctness criterion. However, it is still valid as a definition of truth.

\begin{definition}[Euler-Poincaré invariant]
Let $(V, E)$ be a bipartite multigraph. If $|Cy|$ is the number of minimal cycles and $|CC|$ is the number of connected components, then we have $|V| - |E| + |Cy| - |CC| = 0$.
\end{definition}

Since we are interested in trees, we have $|CC| = 1$ and $|Cy| = 0$. We can divide $V$ into $V_1$ and $V_2$ since it is bipartite. Hence:
\[2(|V_1|+|V_2|-|E|) = 2|V_1|+2|V_2|-2|E| = (2|V_1|-|E|)+(2|V_2|-|E|) = 2.\]

Considering that constellations corresponding to proof-structures and tests are related to partitions, the above equations induce a weight on constellations $\Phi$ defined by $\omega(\Phi) := 2|\Phi|-|\pm\mathtt{Rays}(\Phi)|$. For instance, for a constellation of $\fu \otimes \fu$, we would have a weight of $2 \times 2 - 2 = 2$ and a weight of $2 \times 1 - 2 = 0$ for $\fu \parr \fu$. It is then expected that the sum of the weight of two orthogonal constellations is $2$. This is the case with the two previous dual constellations.

We define the weight of a behaviour as the maximal weight of its constellations and obtain the following weights:
\[\omega(\fu) := 1 \qquad \omega(\mathbf{A} \otimes \mathbf{B}) := \omega(\mathbf{A}) + \omega(\mathbf{B}) \qquad \omega(\mathbf{A} \parr \mathbf{B}) := \omega(\mathbf{A}) + \omega(\mathbf{B}) - 2\]
\[\omega(\mathbf{A}^\bot) := 2-\omega(\mathbf{A}) \qquad \omega(\mathbf{A} \multimap \mathbf{B}) := \omega(\mathbf{B}) - \omega(\mathbf{A})\]

We can now design our notion of truth (but omit the proof that it is invariant by cut-elimination and that $\mathbf{0}$ is false).

\begin{definition}[Visibility]
A constellation or behaviour is \emph{visible} (true) when its weight is a least $0$.
\end{definition}

Visibility is preserved by cut-elimination because when applying a cut on two binary stars representing axioms, we annihilate two rays and two stars are merged into a single star (hence two stars from each constellations are removed but a new one appear):
\[\omega(\exec(\Phi \uplus \Phi')) = 2(|\Phi|+|\Phi'|-2+1)-(|\pm\mathtt{Rays}(\Phi)|+|\pm\mathtt{Rays}(\Phi')|-2)\]
\[2(|\Phi|+|\Phi'|-1)-(|\pm\mathtt{Rays}(\Phi)|+|\pm\mathtt{Rays}(\Phi')|-2)\]
\[= 2|\Phi|+2|\Phi'|-2-|\pm\mathtt{Rays}(\Phi)|-|\pm\mathtt{Rays}(\Phi')|+2\]
\[= 2|\Phi|+2|\Phi'|-|\pm\mathtt{Rays}(\Phi)|-|\pm\mathtt{Rays}(\Phi')|\]
\[= \omega(\Phi \uplus \Phi').\]

We can give a weight to $\wo$ so that it does not make both the behaviours $\mathbf{A}$ and $\mathbf{A} \Rightarrow \mathbf{0}$ visible at the same time. We obtain the following definitions:
\[\omega(\wo) := 0 \qquad \omega(\mathbf{A} \parr \mathbf{B}) := \omega(\mathbf{A}) + \omega(\mathbf{B}) - 2 \text{ (if one of $\mathbf{A}$ or $\mathbf{B}$ does not contain $\wo$)}\]
\[\omega(\mathbf{A} \parr \mathbf{B}) := \omega(\mathbf{A}) + \omega(\mathbf{B}) \text{ (otherwise)}.\]

We have $\omega(\wo \parr \wo) = 0$ and $\omega(\fu \otimes \wo) = 1$ but $\omega(\fu \parr \wo) = -1$ as expected. It is then possible to obtain the following table of Girard's truth where $0$ stands for false (not visible) and $1$ for true (visible), and we ignore the uninteresting truth values:

\begin{center}
\begin{tabular}{|cc||ccc|}
\hline
$\mathbf{A}$ & $\mathbf{B}$ & $\mathbf{A} \otimes \mathbf{B}$ & $\mathbf{A} \parr \mathbf{B}$ & $\mathbf{A}^\bot$ \\
\hline
$1$ & $1$ &  & $0$ & $1$ \\
$0$ & $1$ & $1$ & $0$ & \\
\hline
\end{tabular}
\end{center}
To illustrate a case of this table, if we have $\mathbf{A} := \wo \parr \wo$ (which is not visible since $\omega(\fu \parr \fu) = 0+0-2 = -2$) and $\mathbf{B} := \fu \otimes \fu$ (which is visible since $\omega(\fu \otimes \fu) = 1+1 = 2$), then we have $\omega(\mathbf{A} \otimes \mathbf{B}) = -2+2 = 0 \geq 0$ hence $\mathbf{A} \otimes \mathbf{B}$ is visible.

\emph{``A definite jailbreak from tarskism... and any sort of semantics.''} would say Girard.

\section{Thoughts about the future of logic}

\paragraph{Logic engineering} This expression already appeared in the literature \cite{huth2004logic} but I believe that it can take a new meaning from a transcendental syntax. In programming, the idea of testing and specification is essential. We want programs to have a certain behaviour described by a specification so that the programs does not crash or do unexpected things. These specifications are usually defined relatively to a logic, hence the conception of logic is especially important. For instance, in \emph{model checking} \cite{huth2004logic}, LTL formulas are used to speak about temporal properties about a model of computation (usually an automata or a transition system) such as the fact that a property will eventually happen or that it always holds. To tell if an automata $A$ satisfies an LTL formula $\varphi$, we construct an automata $A_{\lnot\varphi}$ from $\lnot\varphi$ and check if the product $A \times A_{\lnot\varphi}$ recognises the empty language. It is similar to how tests for proof-structures are sort of proof of the negation of a formula. We obtain the following table:

\begin{center}
\begin{tabular}{c||c|c|c|c}
\hline
 & Tested & Test & Specification & Interaction \\
\hline
Model checking & Automata $A$ & $A_{\lnot\varphi}$ & $\varphi$ (LTL) & $\mathcal{L}(A \times A_{\lnot\varphi})$ \\
\hline
Proof-nets & Vehicle $\Phi_\mathcal{S}^\mathrm{ax}$ & $\mathtt{Tests}(\vdash\Gamma)$ & $\vdash\Gamma$ (MLL) & $\exec(\Phi_\mathcal{S}^\mathrm{ax} \uplus \Phi_\mathrm{test})$ \\
\end{tabular}
\end{center}

It may be possible from the transcendental syntax to consider a model checker (or other verification tools) which is independent of a specific logic but in which logics can be defined as sort of modules/libraries. Moreover, it may open ideas of high-order model checking for lambda-calculus since the lambda-calculus can be encoded with proof-structures.

\paragraph{Epidictic architectures} Something important which is left by Girard is \emph{epidictic}. When we quantify in second logic (over a predicate for instance), there are allowed and disallowed properties. But who decide what is licit or not? There is actually something external (what Girard calls a \emph{system} or \emph{epidictic architecture}) which structures the space of all formulas. This is something which appears for instance in type theory with dependent types or in statements such as ``$A$ is a type'' or ``if $A$ and $B$ are types then $A \times B$ is a type''. But how should this be handled? Should we consider other tests constraining the possible tests for quantified formulas/witnesses? More than rules on syntactic labels, our approach should have a computational content, something like a regulation by interaction with some tests. In the case of Girard's first-order, no structuration nor external control is necessary.

\paragraph{Open systems} Without considering systems, the transcendental syntax is naturally \emph{system-free}, in the sense that formulas are characterised by finite sets of tests which are freely chosen. In particular, multiple connectives can live together (such as the difference sorts of exponentials $\ltimes, \otimes, \uppar, \downtens, \oc, \wn$ etc). It is also possible to design any kind of connectives either by finite tests (l'usine) or by interactive typing (l'usage). Even when considering an epidictic architecture (a system), it is possible to change the system as if we were loading a module or library in programming. In particular, is it possible to design a very generic model checker or proof assistant which is independent of a logic but can load any logic? The logic would then be external rather than internal and hard-wired. Trust in the proof assistant or the model checker would only rely on a unification algorithm and tests designed and written by the user.

\paragraph{Logico-functional space} Surprisingly, the encoding of $\lambda$-calculus in stellar resolution shows that logic programs and functional programs live in the same space. Logic programs can be used to represent the structure of $\lambda$-terms (following Regnier's encoding with proof-nets \cite{regnierPHD}) by independent concurrent agents expressed in a language which shares the same dynamics as first-order resolution, itself used in logic programming. This also has similarities with Saurin's use of proof-search in the context of ludics \cite{saurin2008towards}. In the same way, using the stellar resolution, it may be possible to do proof-search on $\lambda$-terms and proof-structures or proof-nets (by adding correctness tests as an additional constraint).

\paragraph{Towards a materialistic logic?} What follows is a possibly exaggerated discussion about how logic could actually be more than our current logic. Something I find quite remarkable is that the stellar resolution places the link between computation and logic in the broad world of complex systems. We compute by non-deterministic local interactions between independent entities with a notion of information propagation. A logic emerges from the behaviour of constellations. We could also expect relationships with sequential dynamical systems and graph dynamical systems.

At the time of the document, I believe than no one truly understands the nature of logic yet. The Geometry of Interaction and the transcendental syntax showed that we can speak about (linear) logic with topology/geometry/dynamics by considering the locations of entities, their links, cycles, interactions, paths etc (see Seiller's graphings \cite{seiller2017interaction}). These are rather tangible things reminiscent of interaction in the biological/physical world (for instance chemical reaction networks \cite{jost2019hypergraph}).

I wonder if it is possible to consider a setting even more general than stellar resolution, for instance a dynamical system for which attractors would have a logical meaning. It may be possible that what we know about logic is only a particular case of a more general \emph{theory of interaction} as mentioned several times already \cite{abramsky2008information, goi0} or a \emph{logic of things} which speaks about interactions in general and not only interaction in the language. But is all sorts of interaction expressible in the language? In particular, mathematics are able to describe physical phenomena only by language and interaction within language.

Indeed, we must take several additional things into account. First, logic has a subjective side because we look at computation from a specific point of view and a choice of what a sound/successful computation means, which depends on the use of objects and our own perception/goals. It probably makes purely mathematical techniques insufficient (unless we speak about the objective part of logic, i.e. computation). A more conceptual/philosophical study and connexions with other fields (typically physics and biology) may be useful and seem possible. Moreover, we can also wonder if logic exists beyond the computable. Does this even make sense?

Finally, I am curious about whether all of that has something to say at all about computational complexity (some hints seem to exists in Seiller's works \cite{seiller2018interaction}). Due to the barriers of problems such as the separations of classes, it may be necessary to look for new definitions of logic and computation.

\bibliography{references}

\begin{appendices}
\section{Not a pure waste of paper (light version)}

In this section, I would like to clarify Girard's terminology. Please read it after being a bit familiar with the content of this paper.

\paragraph{Analytic} In reference to Kant's epistemology. A space of meaningless and computational entities. A ground for logic where objects are expressed before obtaining a logical meaning. A good analytic space should be as natural and simple as possible (typically including independent agents with a local and concurrent interaction so that it is possible to get rid of external control).

\paragraph{Anarchy} Refers to Girard's first-order logic where logical operations are local with no reference to the shape of all possible formulas (no genericity). There is no need of external control providing the logic is limited to some connectives.

\paragraph{Animist} In reference to Japanese traditional beliefs where every entities (including lifeless ones) can have both a subjective part (a spirit inhabiting it) and an objective part (its physical materialisation). In the transcendental syntax, we distinguish \emph{objective} and \emph{subjective} rays. An \emph{animist} star mixes both objective and subjective ones. We usually consider correct a constellation with no animist star (hence, having a clear separation between object and subject).

\paragraph{Apodictic} State of absolute certainty where only finitely many tests with a tractable testing are sufficient for type checking with no appearance of doubts. It happens for Girard's first-order logic but not in general.

\paragraph{Axiom} Group of physical locations (usually two) with are related so that they are updated simultaneously. They are represented with binary polarised stars.

\paragraph{Behaviour} A set of constellations $\mathbf{A}$ with satisfies the biorthogonal closure $\mathbf{A} = \mathbf{A}^{\bot\bot}$ ensuring that $\mathbf{A}$ is characterised by a (potentially infinite) set of tests and that it is theoretically \emph{testable}.

\paragraph{Bureaucracy} Refers to unecessary structures or steps of computation which prevent us from accessing a more explicit version of an entity. For instance, in the sequent calculus, the order of application of rules is usually irrelevant and disappears in proof-nets.

\paragraph{Certainty} State in which the connexion (testing) between answer (computational entity) and a question (formula/behaviour) is purely analytic. In particular, finite testing is able to perfectly ensure the use/interaction of computational entities.

\paragraph{Constat} Refers to irreducible objects (for instance, the normal form of a $\lambda$-term). Also refers to dynamic objects which are seen as static (for instance a program seen as code which can be manipulated as a single entity).

\paragraph{Cut} Link connecting two physical locations which can be seen as an adapter (for instance HDMI/VGA) ensuring that two constellations can be connected by some rays. They are represented with binary stars with a polarity opposite to axioms.

\paragraph{Cut-elimination} Resolution of addresses applied on cuts in order to identify some locations. It defines the dynamics happening in logic, allowing to go from an a reducible object to an irreducible one. This is an access to explicit knowledge by computation. It can also be understood as progressing in an explanation.

\paragraph{Derealism} Approach appearing in Girard's second-order logic where, instead of testing a vehicle (computational entity) against a test (correctness criterion), another test called \emph{mould} for existential witnesses is attached to the vehicle. Hence, a part of subjectivity is included in the object (tested).

\paragraph{Doubt} State in which it is not possible to fully ensure a correct use/interaction of computational objects by finite and tractable testing.

\paragraph{Dichology} See \emph{Behaviour}.

\paragraph{Epidictic} Refers to reasonable but not absolute certainty occurring in logic. It happens in mathematics but also in Girard's \emph{Derealism} when full certainty is not possible because of the presence of correctness test given with the vehicle (hence on the side of tested entities).

\paragraph{Epidictic architecture} In Girard's second-order logic, quantification is general and not limited to a system shaping formulas. Hence, a structuration of the space of all possible formulas is needed for generic statements.

\paragraph{Epistate} Computational object generalising proofs and tests asserting their logical correctness. Another name for \emph{pre-proof} (an object not yet a correct proof but which can be identified as such by some criterion).

\paragraph{Epure} Correct proof in second-order logic. In reference to a French word representing a drawing with several views of an object. These several views are represented by the different tests included with the existential witness provided in a second-order proof. Epures are represented with a constellation having a clear separation between objective and subjective rays (hence no animist star).

\paragraph{Essentialism} Philosophy in which the essence of objects comes before their existence. It corresponds to Church typing where any computational entities only exist attached with a syntactic labels called his \emph{type} and which forbid some interactions considered wrong. In the transcendental syntax, it is updated so to get rid of external semantics and corresponds to typing by finite testing where arbitrary tests define the type of computational objects.

\paragraph{Existentialism} Philosophy in which the essence of objects is defined by their existence. It corresponds to Curry typing where computational entities first exist independently then it is possible, afterwards, to attach a syntactic label on it describing its computational behaviour. In the transcendental syntax, it refers to interactive typing where types correspond to classification of computational objects according to how they interact with each other.

\paragraph{Factory} Typing by finite and tractable testing.

\paragraph{First-order logic} Fragment of logic where no genericity exist and reasoning is purely intrinsic with no need of external control. Logical rules are, for instance, physical moving specific entities such as reuniting two formulas (actually just physical locations) together for the conjunction.

\paragraph{Format} In the transcendental syntax, logic is seen as a way to format computational entities. In particular, there are two way of formatting (or giving meaning). Either by finite testing (see \emph{Factory}) or by interaction (see \emph{Use}).

\paragraph{Formula} Syntactic label materialising an assertion which usually formalises a question or a property. It can be understood either as a constraint or a synthetic description.

\paragraph{Hidden files} Often wrong or ill-behaving entities not considered in traditional proof theory and whose existence is computationally relevant in logic. For instance, the formula $0$ in linear logic is considered as a contradiction with no proof and thus corresponds to an empty set. In the transcendental syntax, the corresponding behaviour $\mathbf{0}$ is inhabited by constellations with none of them considered correct. However, they can still interact with other constellations and have a computational meaning.

\paragraph{Interaction} Phenomena occurring when opposing two compatible objects. In the transcendental syntax, this is represented by the execution of the union of two constellations.

\paragraph{Locativity} Approach taken in Ludics and the latest articles of GoI where the physical location, referred to by an address is given a great importance. For instance, the formulas in a sequent calculus proof are inessential labels which can be replaced by addresses (usually natural numbers). A proof becomes a manipulation of addresses.

\paragraph{Monism} Approach in which every entities is represented as objects of the same kind. For instance, proof-nets and their tests become constellations but so are circuits and their semantics, automata and words, tile systems and Turing machines. It is the opposite of \emph{dualism} where we distinguish objects of two kinds (for instance program vs environment, syntax vs semantics etc).

\paragraph{Morphologism} Girard's paradigm for logic where what matters is the shape of objects and their interactions.

\paragraph{Objective rays} Usual rays which is used in Girard's first-order logic. They are used to represent proof-structures.

\paragraph{Ordeal} See \emph{Test}.

\paragraph{Orthogonality} Binary symmetric relationship formalising the compatibility of two computational entities. In the transcendental syntax, it also formalises the idea of ``passing a test''. We can design any orthogonality relationship we want and it leads to different computational analyses or models of logic.

\paragraph{Performance} Procedure leading a reducible computational object to an irreducible one. It is essential in the theory of computation since reducible objects cannot be reduced to their results because of the computational indecidability (for instance, the halting problem).

\paragraph{Predicate calculus} A fragment of Girard's second-order logic where there are two sorts of quantification: one over predicates and another one over individuals. Both are encoded as MLL behaviours. Hence, it corresponds to second-order logic with an epidictic architecture limited to multiplicative linear logic. Equality becomes linear equivalence between MLL formulas instead of mere predicate.

\paragraph{Proof} Object formalising an answer to a formula. It can take several forms (tree, graph, number etc) but usually depends on the requirement of the formula and what is considered an answer to the formula. In logic, we are usually interested in the adequacy between proofs and formulas (did we correctly answer the question?).

\paragraph{Realism} Belief in which the meaning of an entity is located in an external space of \emph{reality}, what typically happens in denotational semantics where syntactic objects are interpreted in another (mathematical) space. For instance, $x = x$ because the two occurrences of $x$ refer to the same reality. In the case of logic, logical systems can be mistakenly considered as correct representations/materialisations of reality.

\paragraph{Second-order logic} A fragment of logic which uses generic reasoning. For instance, the rule of the modus ponens (from $A$ and $A \Rightarrow B$ follows $B$) works for any formula $A$ and $B$. Girard claims that a large part of traditional logic is actually second-order (propositional logic and first-order logic in particular). In the transcendental syntax, second-order logic has a different meaning. It corresponds to a logic where quantification is unlimited and does not presuppose how the space of all formulas is shaped. It needs a specific external architecture.

\paragraph{Semantics} System in which we associate a meaning to (often computational) objects. For instance, denotational semantic associates a mathematical object to proofs. Such objects represent the meaning of the proof, which is purely syntactic.

\paragraph{Scientism} Refers (usually negatively) to a belief in which science and more especially logic can explain everything. In particular, certainty is absolute and logic has the role of accessing knowledge and solving philosophical problems. It is then sufficient to compute an answer to access knowledge.

\paragraph{Stars and constellations} Computational objects used in the transcendental syntax. The terminology comes from the fact that the stars are objects with can be connected to other ones along its branches. It could also be called atom or molecule but this would be less poetic.

\paragraph{Subjective rays} The objective rays are the ordinary rays. We then arbitrarily distinguish another class of \emph{subjective} rays which are used to discriminate some constellations so that it is possible to define notions of truth or logical correctness.

\paragraph{Synthetic} In reference to Kant's epistemology. Space where a meaning is given to meaningless computational objects. In the transcendental syntax, it is represented by typing by finite testing and interactive typing. It is what allows us to design formulas/types or questions/properties.

\paragraph{Test} Way to assert that an object has a specific property or computational behaviour we expected.

\paragraph{Type} Same as formula by the Curry-Howard correspondence but more used in the context of programming and the theory of computation where types (sometimes called specifications) are used as labels preventing ill behaviours or unwanted situations to happen during computation.

\paragraph{Usage} See \emph{Use}.

\paragraph{Use} Refers to how logical definitions can be used in practice. In the transcendental syntax it appears in interactive typing where constellations are given a meaning depending on how can they can be used (how they can interact with other constellations). According to Girard's, Gödel's incompleteness theorems state that tests cannot always guarantee the use we wish for.

\paragraph{Usine} See \emph{Factory}.

\paragraph{Vehicle} The computational content of a proof. In the theory of proof-nets, it corresponds to the set of axioms of a proof-structure. It is the part which interacts with tests. The tests will then ensure that the vehicle has a specific computational behaviours when interacting with other objects (the vehicle of another proof). For instance, MLL+MIX correctness guarantees strong normalisation of cut-elimination.

\paragraph{Wire} Analogy used by Girard in order to talk about logic in a more concrete and down-to-earth way. In particular, the theory of proof-nets sees proofs as wiring of formulas and the cut-elimination as a resolution of wiring. The transcendental syntax extends this analogy since, in the stellar resolution, wires are terms $c(x)$ which can be divided into infinitely many sub-wires, for instance $c(\mathtt{l} \cdot x)$ and $c(\mathtt{r} \cdot x)$, while keeping finite objects.

\end{appendices}

\end{document}